\documentclass[structabstract]{aa}
\usepackage{txfonts,epsfig,graphicx,natbib,url,twoopt}
\usepackage{url}
\usepackage{booktabs}		
\usepackage{subfigure}
\usepackage{natbib}  
\usepackage[breaklinks=true]{hyperref} 
\usepackage[labelfont=footnotesize,textfont=footnotesize]{caption}  
\usepackage{float}
\bibpunct{(}{)}{;}{a}{}{,}    
\newcommandtwoopt{\citeads}[3][][]{\href{http://adsabs.harvard.edu/abs/#3}%
                                        {\citealp[#1][#2]{#3}}}
\newcommandtwoopt{\citepads}[3][][]{\href{http://adsabs.harvard.edu/abs/#3}%
                                         {\citep[#1][#2]{#3}}}
\newcommandtwoopt{\citetads}[3][][]{\href{http://adsabs.harvard.edu/abs/#3}%
                                         {\citet[#1][#2]{#3}}}
\newcommandtwoopt{\citeyearrads}%
  [3][][]{\href{http://adsabs.harvard.edu/abs/#3}%
                                         {\citeyear[#1][#2]{#3}}}
\usepackage{caption}	      
\newcommand{\hi}{{H\sc{i}~}}

\begin{document}  
\title{H{\sc i} observations of three compact high-velocity clouds around the Milky Way}

   \author{S. Faridani \inst{1}, L. Fl\"oer \inst{1}, J. Kerp \inst{1}, T. Westmeier \inst{2}
          }

   \institute{$^1$Argelander-Institut f\"ur Astronomie,
             Auf dem H\"ugel 71, D-53121 Bonn, Germany\\
                  $^2$International Centre for Radio Astronomy Research (ICRAR), The University of Western Australia, 35 Stirling Highway, Crawley WA 6009, Australia\\
                   }
\date{Received 12 September 2013 / Accepted 20 January 2014}

\abstract
{We present deep \ion{H}{i} observations of three compact high-velocity clouds (CHVCs).}  
{The main goal is to study their diffuse warm gas and compact cold cores. We use both low- and high-resolution data obtained with the 100 m Effelsberg telescope and the Westerbork Synthesis Radio Telescope (WSRT). The combination is essential in order to study the morphological properties of the clouds since the single-dish telescope lacks a sufficient angular resolution while the interferometer misses a large portion of the diffuse gas.}
{Here single-dish and interferometer data are combined in the image domain with a new combination pipeline. The combination makes it possible to examine interactions between the clouds and their surrounding environment in great detail.}
{The apparent difference between single-dish and radio interferometer total flux densities shows that the CHVCs contain a considerable amount of diffuse gas with low brightness temperatures. A Gaussian decomposition indicates that the clouds consist predominantly of warm gas.}
{}

\titlerunning{HI observations of three CHVCs}
\maketitle

\section{Introduction}\label{sec:introduction}

High-velocity clouds (HVCs) are neutral atomic hydrogen (\ion{H}{i}) clouds with radial velocities incompatible with simple Galactic rotation models \citep{1997ARA&A..35..217W}.
After their initial discovery by \citet{1963CRAS..257.1661M}, they have been observed all around the Milky Way.
\citet{1999ApJ...514..818B} suggested that HVCs are distributed throughout the Local Group, but this theory has been disproven several times by searching for HVCs around other galaxies: all these observations have either failed to find any HVCs (\citet{2004ApJ...610L..17P, 2007ApJ...662..959P}) or located them very close to their host galaxies \citep{2004ApJ...601L..39T}. High-velocity clouds are found either as compact and isolated objects or as part of large complexes \citep{1997ARA&A..35..217W}.

Recent studies propose three main hypotheses for the origin of HVCs \citep{2004ASSL..312..341B}. First, HVCs consist of primordial gas that is accreted onto galaxies. It can be primordial gas flows from the filaments or have its origin in gaseous dark matter haloes. Second, HVCs originate from the tidal and ram-pressure interaction of the Milky Way Galaxy with dwarf galaxies. Third, HVCs have been formed as the result of a galactic fountain, i.e., by gas flows driven by supernovae. The most serious problem is the limited information on the HVC distances \citep{2000ASPC..218..407V, 2004ASSL..312..195V, 2005A&A...440..775K, 2008ASPC..393..179B}.

Investigations of the physical conditions of HVCs indicate that some show signs of interaction with the Galactic halo gas \citep{2000A&A...357..120B, 2001A&A...370L..26B, 2005A&A...432..937W, 2011MNRAS.418.1575P, 2012A&A...547A..12V}. 

The aim of our study is to investigate the morphological properties of compact high-velocity clouds (CHVCs) as well as their radial velocity and line width. In the past, the existence of two different gradients either in density \citep{2011MNRAS.418.1575P} or velocity \citep{2000A&A...357..120B} were required within the cloud for the designation head-tail (HT). The objects studied in the present paper appear to conform to both these criteria. The appearance of the HT structure is interpreted as follows. As the cloud moves through the hot and ionized halo, part of the cloud is compressed. This higher density area constitutes the head of the cloud. Part of the gas is stripped off the cloud to form a less dense and thin tail structure that follows the head with velocities lower than those of the cloud bulk motion \citep{2002A&A...391..713K, 2009ApJ...698.1485H}.

The cloud HVC\,125+41--207 is prototypical for interacting HVCs denoted accordingly as head-tail (HT) HVCs. \citet{2000A&A...354..853B} used the Westerbork Synthesis Radio Telescope (WSRT) to study the small--scale structure of compact HVCs (CHVCs). Towards HVC\,125+41--207, they discovered an extremely narrow \hi line unresolved by their spectral resolution of $2.47\,{\rm km\,s^{-1}}$. The peak brightness temperature of this line is exceptionally high with $T_{\rm B} = 75$\,K indicating the existence of a cold and dense cold neutral medium (CNM) core. Their maps reveal a highly structured \hi CNM distribution. \citet{2000A&A...357..120B} used Effelsberg observations of HVC\,125+41--207 to deduce that this dense core is embedded within a warm neutral medium (WNM) envelope. Moreover, they found evidence that the WNM shows signs of deceleration not only towards the cloud's tail but also along the rims. Proportional to this deceleration of the radial velocity component, the WNM gets warmer (see their Fig.\,2). Only the combined analysis of the WSRT and Effelsberg data allowed the deduction of such a homogenous view of HVC\,125+41--207 as an interacting CHVC.

To investigate the gaseous structures of HVCs and their dynamics accurately, the combination of radio interferometric and single-dish data is essential. Radio interferometers are insensitive to structures on the scale of tens of arc minutes and beyond, the WNM. Single-dish observations are unable to resolve the small-scale structure of the CNM. The combination of \ion{H}{i} single-dish and interferometric observations provides the possibility of studying HVCs in the necessary detail.

In this paper, we present high-resolution \ion{H}{i} observations of three CHVCs, using the 100 m Effelsberg telescope and the WSRT.
Section\,\ref{sec:observations} presents the details of the data, and the \ion{H}{i} observations are introduced. Section\,\ref{subsec:method} deals with the method that has been used to combine these two data sets.
Section\,\ref{sec:CHVCs} discusses the physical and morphological properties of the clouds and the results of the Gaussian decomposition of the integrated spectral lines. Section\,\ref{sec:conclusions} summarizes our findings and gives an outlook on future work.

\section{Observations and data} \label{sec:observations}

\begin{figure}[ht]
\centering
\subfigure[CHVC\,070+51-150]{\includegraphics[width=0.51\textwidth]{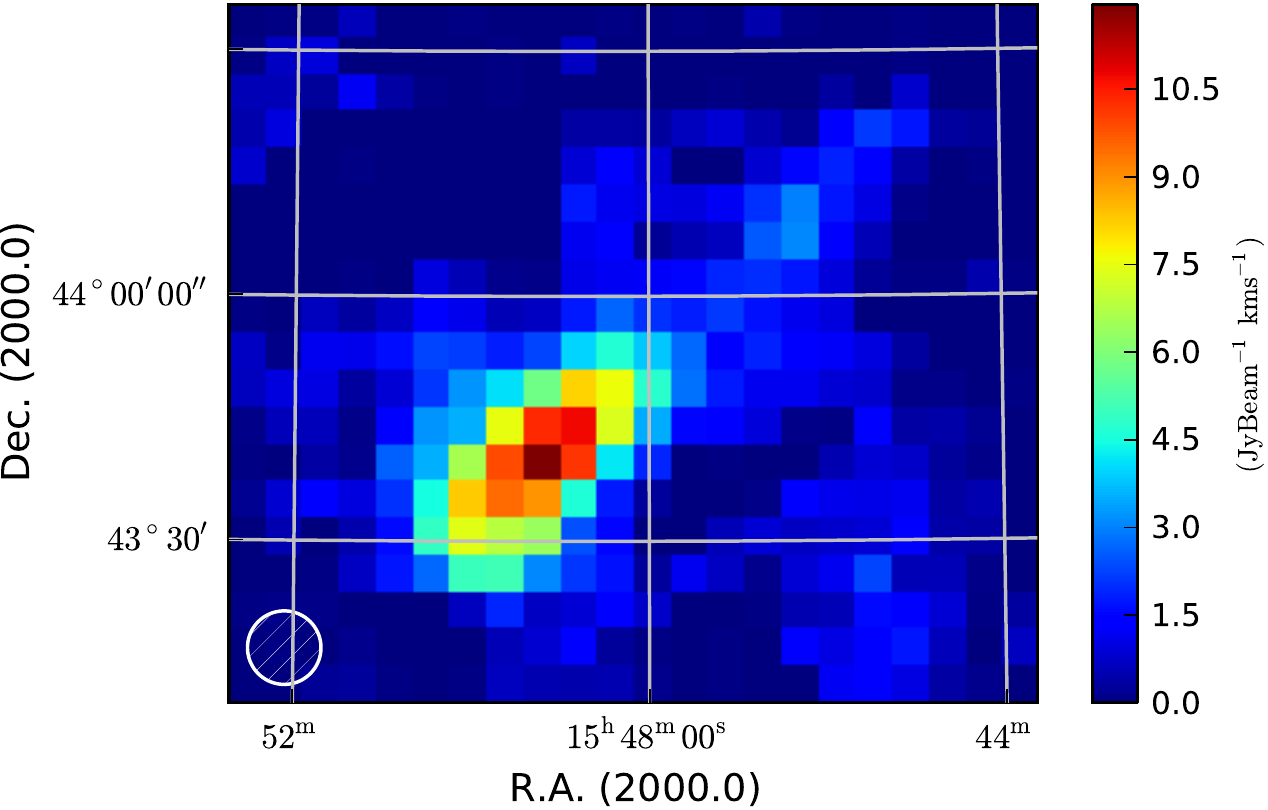}}
\subfigure[CHVC\,108-21-390]{\includegraphics[width=0.51\textwidth]{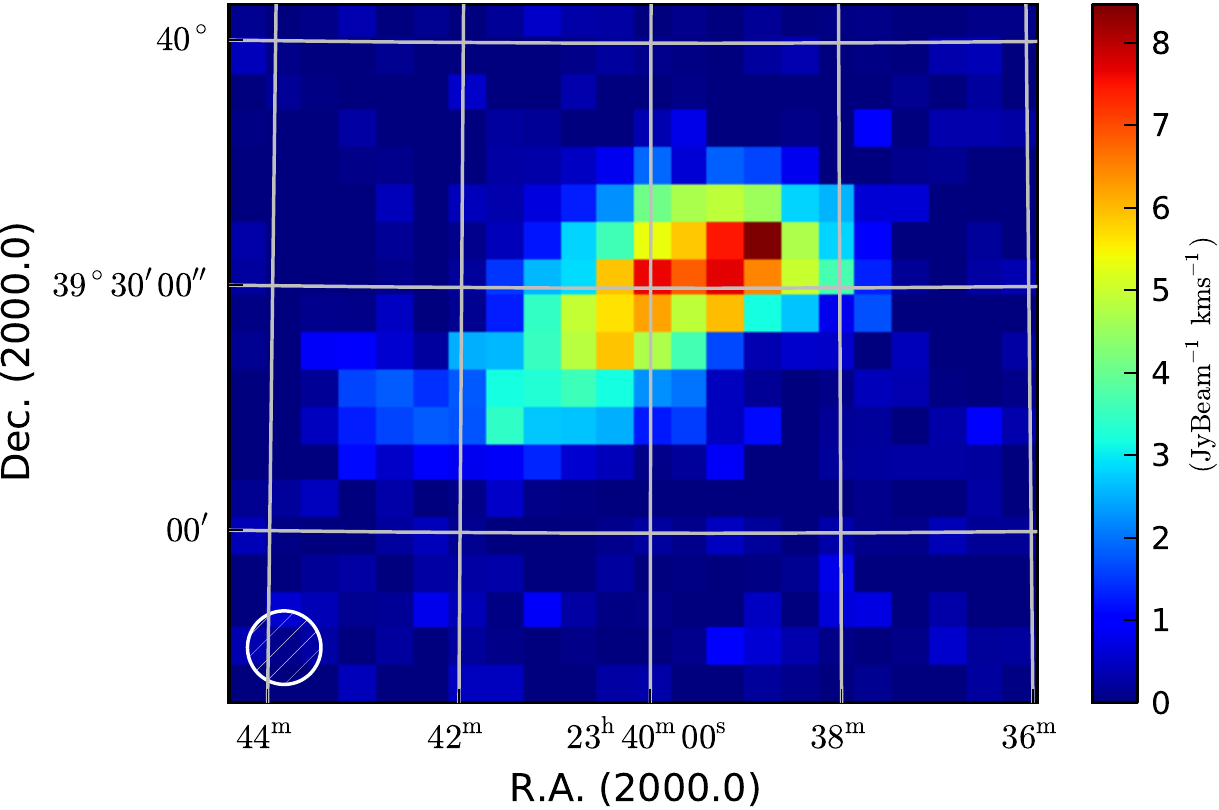}}
\subfigure[CHVC\,162+03-186]{\includegraphics[width=0.51\textwidth]{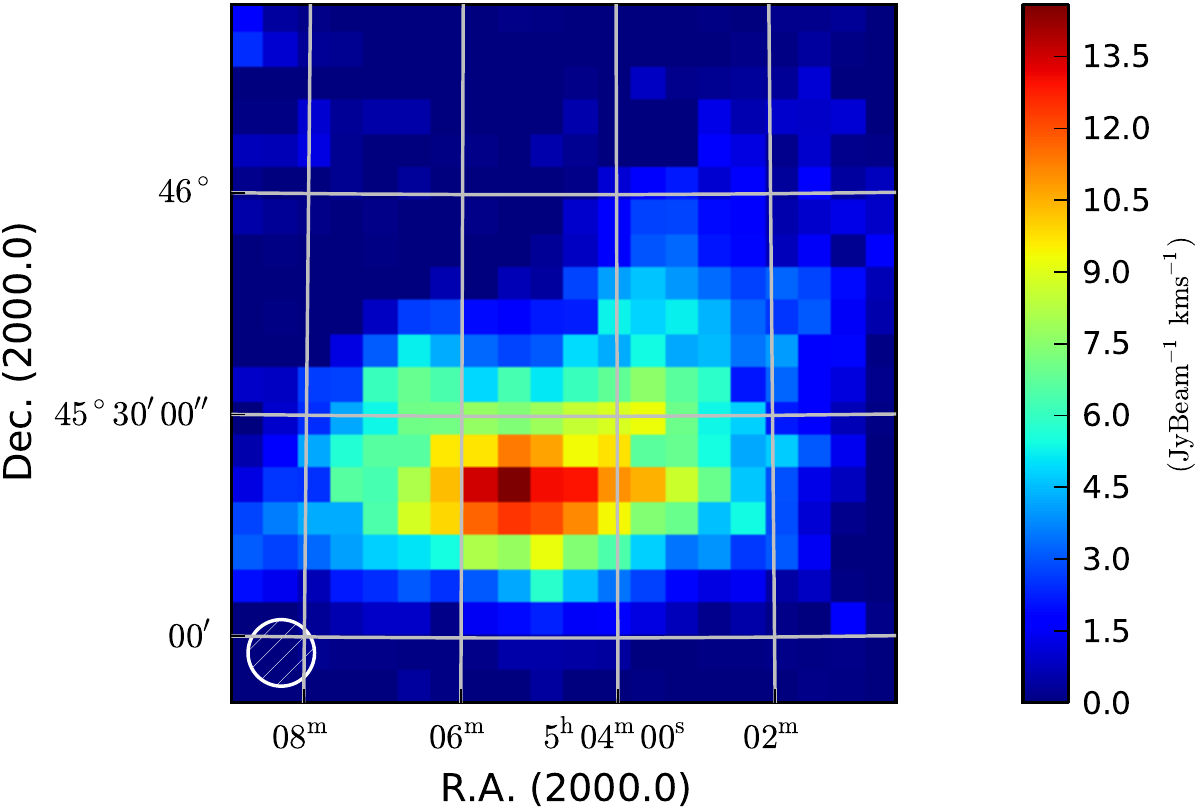}}
\caption{CHVC flux-density maps as observed with the Effelsberg 100 m telescope. Figure (a) presents CHVC\,070+51-150, (b) CHVC\,108-21-390, and (c) CHVC\,162+03-186.}
\label{fig:LR}
\end{figure}

All three CHVCs have non-symmetric, and specifically non-spherical shapes (see Fig.\,\ref{fig:LR}). The complex morphology of their appearance as seen with a single-dish telescope is elongated in their \hi distribution; this has been interpreted as clear sign of a distortion by ram pressure of an ambient medium.

The single-dish observations of the three CHVCs were carried out in January 2000 with the 100 m Effelsberg telescope. The half-power beamwidth (HPBW) of the Effelsberg data is $ \approx\,9'$ (see Table.\,\ref{tb:obs_par} for observational details). Figure\,\ref{fig:LR} shows the Effelsberg integrated flux maps of the clouds. All CHVCs exhibit a HT morphology with a pronounced core. To resolve these cores, radio interferometric observations were performed with the WSRT. Furthermore, the three clouds have declinations of about $+40^{\circ}$ allowing for a complete 12-hour coverage with the WSRT. The combination of both low- and high-resolution data makes it possible to investigate the warm diffuse gas as well as the compact and cold structures in detail.

The interferometric observations of the three CHVCs were carried out in November 2004 and May 2005 in the maxi-short configuration with the WSRT. This setup provides good sensitivity on the short baselines as well as on a long baseline of $2.7~\mathrm{km}$. Each CHVC was integrated for $12~\mathrm{h}$ in a single pointing on the sky centered on the maximum integrated intensity from the single-dish data. For each of the two polarization the correlator provided a total of 1024~spectral channels across a bandwidth of $2.5~\mathrm{MHz}$, resulting in an intrinsic velocity resolution of about $0.5~\mathrm{km \, s}^{-1}$.
  
The data were reduced using the Astronomical Image Processing System (AIPS). After reading in and concatenating the visibility data files, we flagged data affected by radio-frequency interference and shadowing as well as potential \hi absorption lines in the calibrators. Next, we carried out the spectral bandpass calibration using one of the primary calibrators, followed by the external gain calibration on the pair of primary calibrators observed at the beginning and end of each observing session.
  
Observations with the WSRT do not normally include gain calibrators at regular time intervals. Instead, the target of interest is tracked continuously for a period of $12~\mathrm{h}$, and gain calibration is then achieved by self-calibrating on background continuum sources throughout the field. We followed this approach and iteratively self-calibrated both the gain phase and amplitude until we were able to reach the theoretical noise limit of the data.

In radio interferometry, the final image is the representation of the sky as it is seen by the array multiplied by the primary beam response of a single antenna. The response of the antenna can be modeled as a Gaussian then divided out of the image. This correction accounts for the lower sensitivity towards the edges of the primary beam. It should always be the last stage of the imaging process after the best-quality image is produced. Performing the correction at early stages leads to incorrect results \citep{1989ASPC....6..213F}. After the primary beam correction one finds enhanced noise at the edges of the image. However the correction is indispensable for the estimation of the \hi total flux density and \hi mass. For the following three CHVCs it was performed using the \texttt{Miriad} task \texttt{Linmos}.

Although the beam size of the single-dish Effelsberg data is constant at 9\,arcmin, the WSRT synthesized beam varies in size according to the different source declinations. Hour angle coverage and visibility weighting schemes can also cause variations in the beam size. In the case of presented CHVCs the applied weighting scheme is the same, although there are variations in the hour angle coverage. Prior to the combination of both data sets, it is essential that they are reprojected onto the same grid. Regridding was performed within \texttt{MIRIAD} with the help of the \texttt{regrid} task. The regridded single-dish cube and the interferometer cube are the inputs of the combination pipeline that was implemented using various \texttt{CASA} tasks.

\begin{table}[ht]
\centering
\tiny{
\begin{tabular}{l c c c c c c }
\toprule
Name                  	 & $\alpha$ (J2000) 	 & $\delta$ (J2000)  & $v_{\mathrm{LSR}}$  & $\Delta v$                \\
                          &                     &                   &                     & Eff/WSRT/comb.            \\
(CHVC $l\pm b$)		     & [hh:mm:ss]       	 & [dd:mm:ss]        &[${\rm km\,s^{-1}}$]  & [$\mathrm{km\,s^{-1}}$]  \\
\midrule
CHVC\,070+51               & 15:49:13            & 43:39:30          & -150                & 5.15/2.57/2.57 \\
CHVC\,108-21               & 23:38:50            & 39:35:28          & -390                & 2.57/2.57/2.57 \\
CHVC\,162+03               & 05:05:20            & 45:20:29          & -186                & 2.57/2.57/2.57 \\
\bottomrule
\end{tabular}
\caption{Observational details of the CHVCs. $\alpha$ is the right ascension, $\delta$ the declination, $v_{\rm LSR}$ the local-standard-of-rest velocity, and $\Delta v$ is the spectral channel resolution of Effelsberg, WSRT, and combined data.}
\label{tb:obs_par}
}
\end{table}

\subsection{Combination of low- and high-resolution data} \label{subsec:method}

Since the shortest baseline an interferometer can cover is determined by the smallest separation of two adjacent dishes, the central part of the uv-plane is never sampled. This problem, known as the short- or zero-spacing problem, leads to an insensitivity of interferometers towards large angular scales. \citet{1991ASPC...19..188S} show that for some synthesis observations a lack of short spacings cannot be tolerated, for example if the emission fills the whole primary beam. The combination of interferometric and single-dish data provides in this case the only solution to the missing short spacings problem.

The combination of single-dish and radio interferometric data sets can be performed either in the image or in the Fourier domain \citep{1991MNRAS.249..722Y, 2009ApJ...705.1395C, 1979A&A....75..251B, 1984ApJ...283..655V}. To achieve this combination, a number of approaches exist, most of which are not straightforward. Supplementary information is sometimes required, such as the visibility distribution or  the dirty images \citep{2009PASJ...61..873K, 2001A&A...365..571W}. In practice it is common to store only the cleaned images as final data products; retrieving the supplementary information can be difficult or even impossible. Data reduction packages, such as \texttt{CASA} or \texttt{Miriad}, provide easy-to-use tasks to perform this combination using only the cleaned interferometric images and single-dish data, but these packages often act as ``black boxes'', making it difficult to evaluate the accuracy and systematic uncertainties inherent to the applied method.
 
Systematic limitations like the edge-effects are significant when Fourier transforming the radio interferometric data to the $uv$ domain, in particular when the gaseous structures are very extended and even exceed the boundaries of the primary beam. For these cases \citet{1999MNRAS.302..417S} introduce a new approach to mitigate some of these problems. Their method has been evaluated with observations performed with the 64 m Parkes telescope and the Australia Telescope Compact Array. They applied the maximum entropy method to combine both data sets in the image domain. Although the combination achieves convincing results regarding the flux consistency between the original single-dish and the combined data, the method is highly non-linear and additional information such as beams and dirty images is required, since the combination occurs during deconvolution \citep{2002ASPC..278..375S}.  

In our analysis both high-resolution (interferometer) and low-resolution (single-dish) data have been merged. We combined the cleaned radio interferometric images and the single-dish data in the image domain using a linear approach. In the first step, differential projection and geometry for single-dish vs. interferometric data were accounted for; then the interferometric data were smoothed to the resolution of the single dish, and then subtracted from the single-dish data. The resulting difference cube contains only diffuse extended \ion{H}{i} structures detectable by the single dish. Finally, this is added to the interferometric data cube. This procedure allows us to recover both the complete flux (single-dish flux) and the highest angular resolution (interferometer). The full method will be explained in a forthcoming paper \citep{ssc}.

\section{Compact high-velocity clouds} \label{sec:CHVCs}

The term CHVC was defined by \citet{2000A&A...354..853B}. The designation CHVC or HVC is an indication of the characteristics of the clouds. High-velocity clouds are associated with more extended complexes, whereas CHVCs are small compact objects, well separated from the complexes. Compact high-velocity clouds usually have angular sizes smaller than $2^{\circ}$, although different isolation criteria also exist. In any case, the velocity of the cloud is a strong argument in favor of its being part of a structure rather than being isolated \citep{2011MNRAS.418.1575P}. 

All three of the CHVCs observed have a rather complex and irregular morphology. Figure\,\ref{fig:LR} shows the Effelsberg observations of the CHVCs where it can be seen that CHVC\,070+51-150 and CHVC\,108-21-390 both exhibit a pronounced head-tail (HT) structure. In the case of CHVC\,162+03-186 we find a so-called bow-shock shaped CHVC \citep{2005A&A...432..937W}. The asymmetric appearance of the three CHVCs suggests a possible interaction with the ambient medium. All radial velocities of the three clouds are negative in LSR (see Table\,\ref{tb:obs_par}). 

\begin{figure*}[ht]
\centering
\subfigure{\includegraphics[scale=0.46]{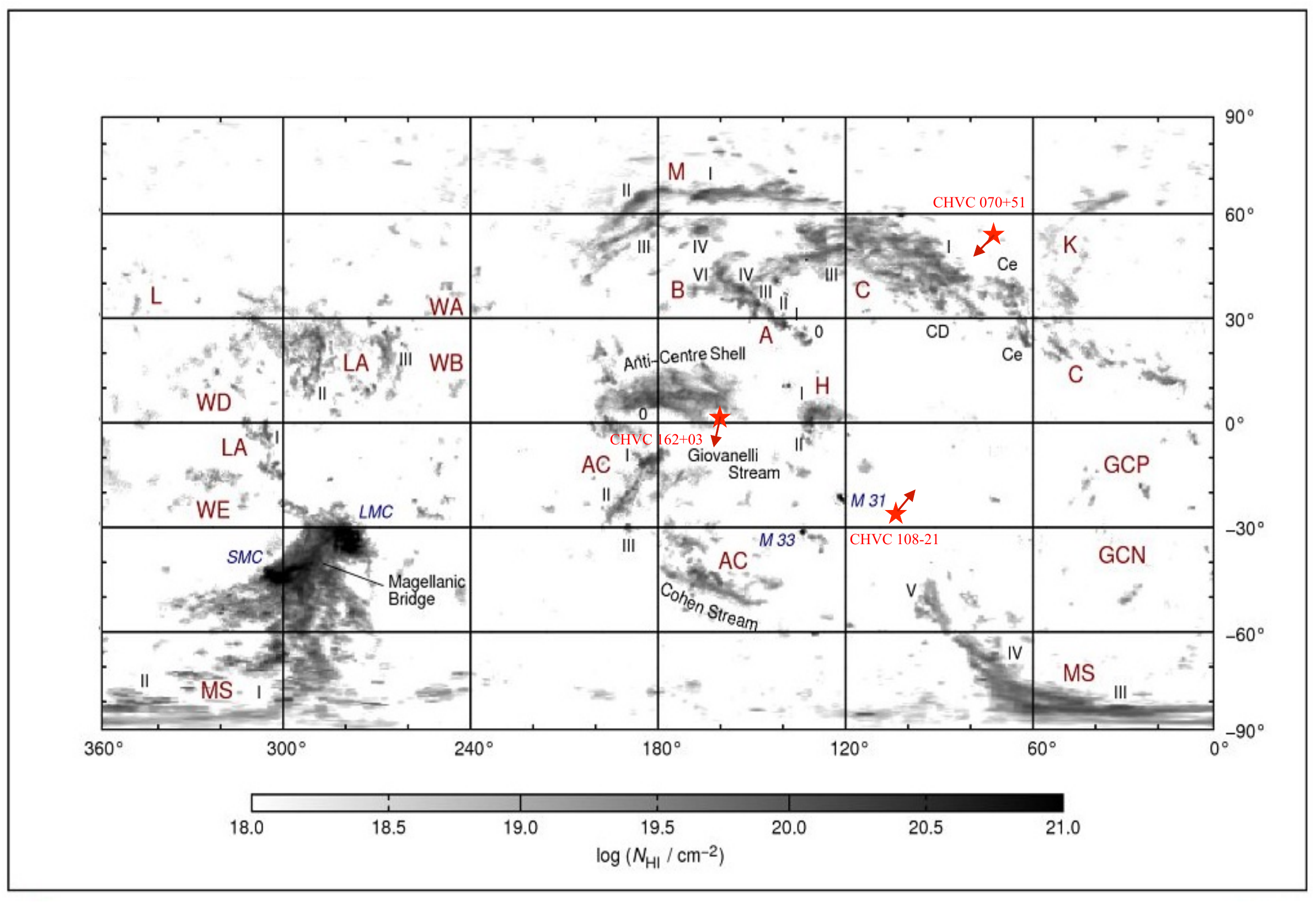}}
\subfigure{\includegraphics[scale=0.50]{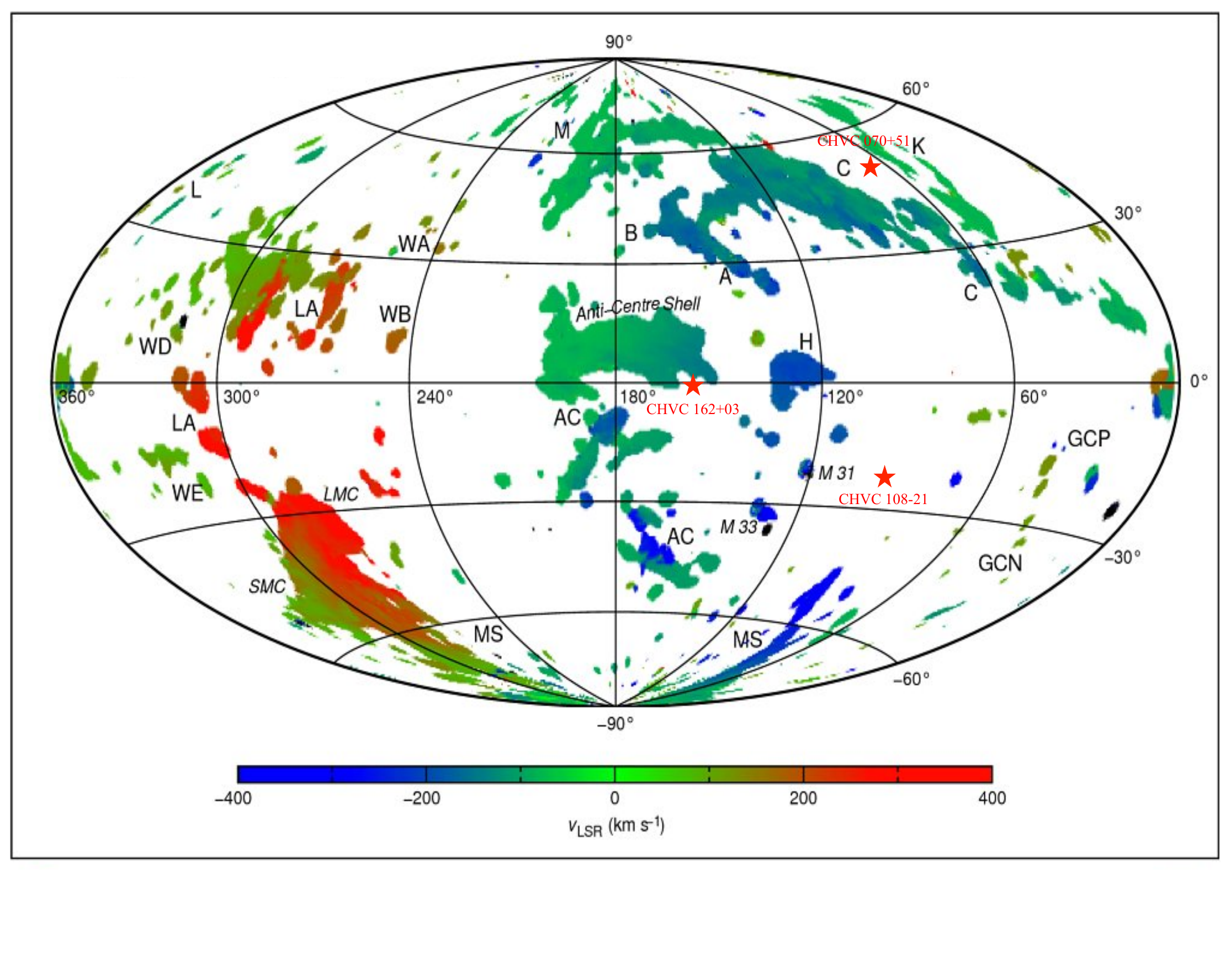}}
\caption{The high-velocity sky based on the Leiden/Argentine/Bonn Survey. The figure at the top shows the location of the three CHVCs. The arrows indicate the projected motion direction on the sky based on the ratio of brightness temperature $T_B$ over kinetic temperature $T_\mathrm{kin}$. The bottom figure shows the velocity distributions of the surrounding structures.}
\label{fig:allsky}
\end{figure*}

In Fig.~\ref{fig:allsky} the positions of the three clouds are marked in the global HVC map generated from the Leiden/Argentine/Bonn (LAB) Galactic \hi survey \citep{2005A&A...440..775K}. Additionally, the velocity distribution map of the high-velocity sky is presented for a comparison of the clouds' velocities with their surrounding structures. The arrows present an estimate of the projected direction of motion on the sky based on the ratio of brightness temperature $T_B$ over kinetic temperature $T_\mathrm{kin}$.

CHVC\,070+51-150 is located between complexes C and K. Complex C is located at a distance $\geq 6$ kpc  \citep{2008ApJ...684..364T, 2004ASSL..312...25W, 2006A&A...455..481K}. \citet{2006A&A...455..481K} present some evidence of the possible existence of multi-component structures. CHVC\,070+51-150 seems to be an isolated compact HVC. Its radial velocity agrees with the presented radial velocities for complex C and K ($\mathrm{V_{LSR}}\, \approx$-150 $\mathrm{km\,s^{-1}}$). CHVC\,162+03-186 is located in the direction of the anti-center shell. It is not fully isolated and is probably associated with the anti-center shell\citep{1986Ap&SS.118..531K}. The measured radial velocity of the cloud agrees with the presented velocities in this region. The measured radial velocity of -186 $\mathrm{km\,s^{-1}}$ is in agreement with the velocity range of the anti-center very high-velocity clouds (ACVHVC) reported by \citet{2006A&A...455..481K}. For this complex both distances and metallicities are unknown. 

CHVC\,108-21-390  is located at the position of the EN complex. In fact, the cloud is listed and shown by \citet{2004A&A...417..421B}. It sits right on top of the Magellanic stream and has the same velocity as the surrounding stream clouds. The notation EN population is derived from the extremely negative (EN) radial velocities measured at this region. According to \citet{2004ASSL..312...25W} the EN complex is located at a distance of approximately 50 kpc. 

The aforementioned distances are useful for driving additional physical properties of the three HVCs. In particular, the distance is essential for the estimation of their \hi masses.

In Figs.\,\ref{fig:chvc-070+51}-\ref{fig:chvc-162+03}, we show the different data sets for each cloud as well as the combined data and a velocity-weighted map. We note that the top-left panel in each figure presents the regridded and reprojected Effelsberg data cropped to the same field of view as in the interferometric and combined maps. It is clear that WSRT single-pointing observations only cover the clouds partially, whereas the Effelsberg observations in Fig.\ref{fig:LR} show the entire extended structure of the clouds down to the observational detection limit. Therefore, the single-dish data cubes are used for the determination of the velocity distribution.

We apply the method of \citet{2012A&A...547A..12V} to derive the projected motion of the clouds in the sky. We calculate the ratio of brightness temperature $T_B$ over kinetic temperature $T_{\rm kin}$. Cold neutral medium structures will show up with high $T_B$ and correspondingly large ratios, while WNM gas is characterized by high $T_{\rm kin}$ values and low ratio values on the map. It is important to mention that the value of $T_{\rm kin}$ is derived from the \hi line width. In this case, the derived value is not the actual kinetic temperature, but only an upper limit, since additional turbulence broadening will have contributed to the line width. Hence, the values of $T_B$/$T_{\rm kin}$ are to be considered lower limits. In the gradient map, the pixel with highest value represents the center of density. Additionally, we determine the intensity maximum in the gradient map. An arrow connects the brightness temperature maximum (arrow head) and the centre of the projected intensity distribution. Its direction indicates the projected motion on the sky based on the ratio of $T_B$ over $T_{\rm kin}$ (Fig. \ref{fig:allsky}, top panel).

\subsection{CHVC\,070+51-150} \label{subsec:070+51}

\begin{figure*}
\centering
\subfigure[CHVC\,070+51-150 - $\mathrm{Effelsberg_{\,reg}}$]{\includegraphics[width=.49\textwidth]{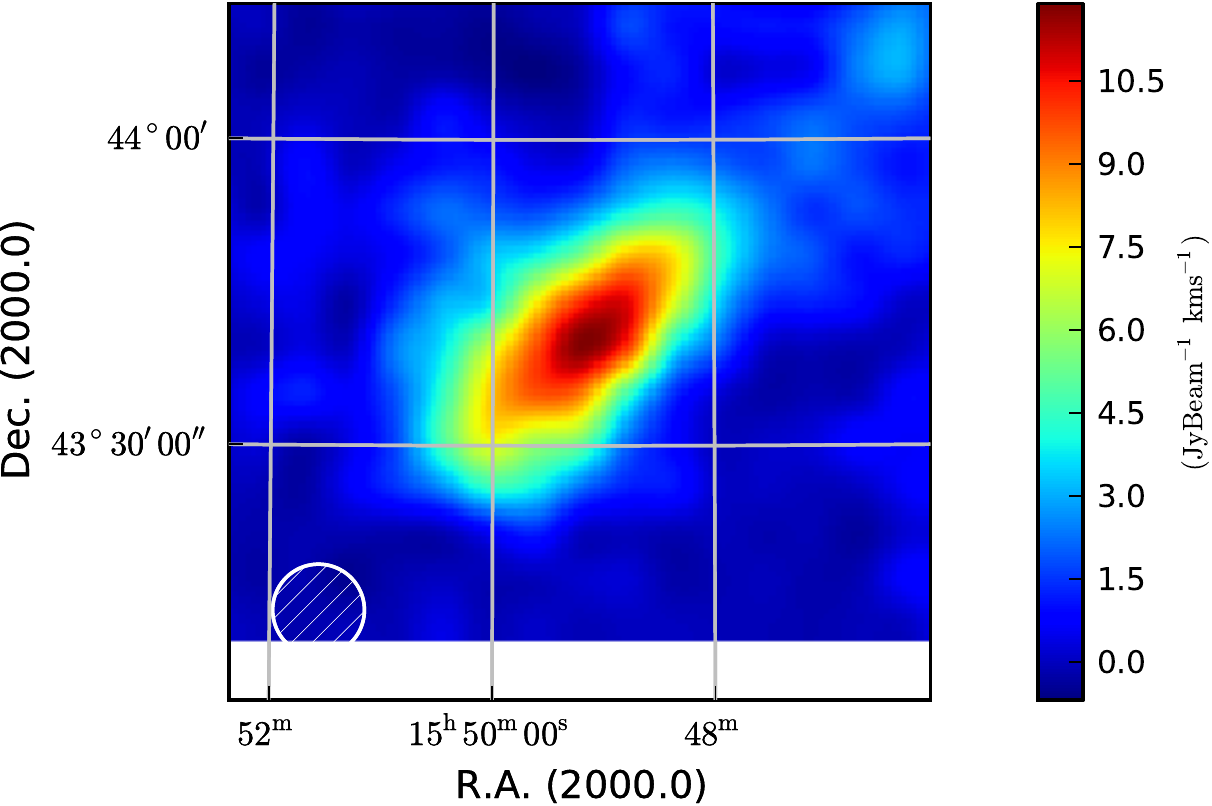}}
\subfigure[CHVC\,070+51-150 - WSRT]{\includegraphics[width=.49\textwidth]{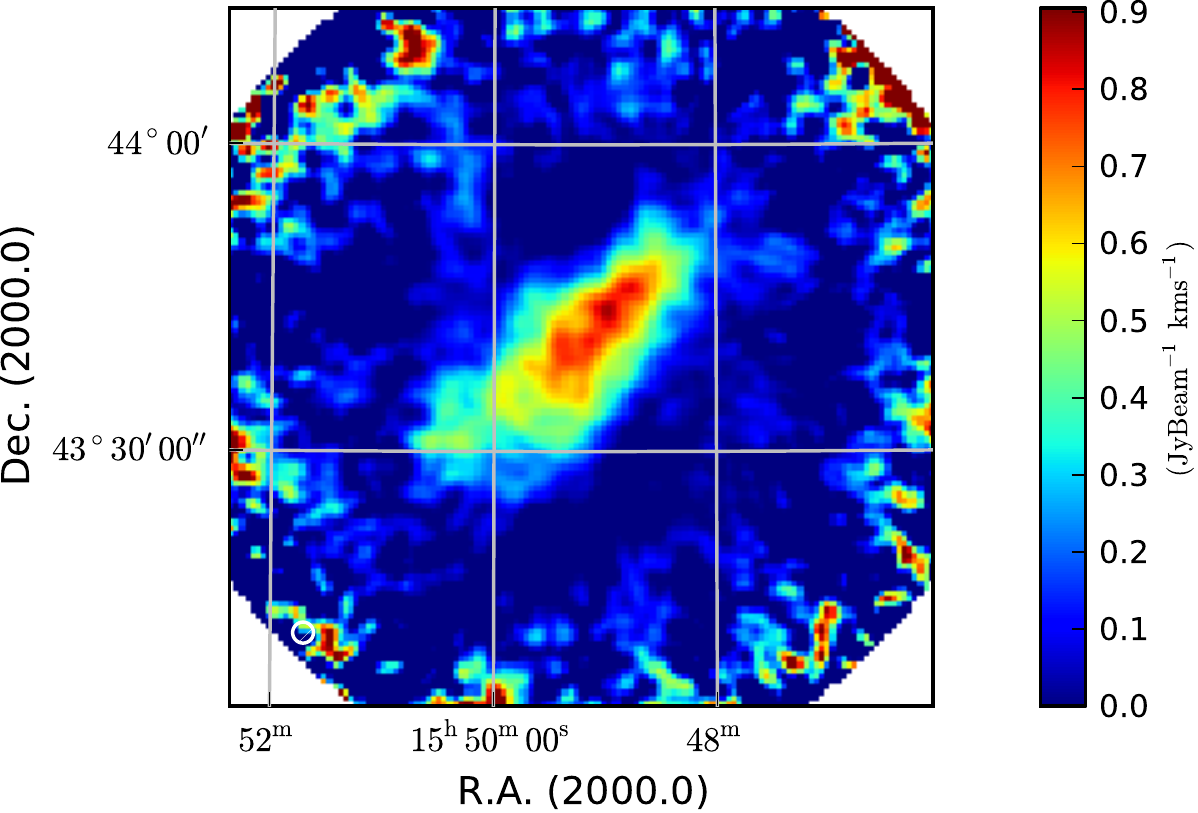}}
\subfigure[CHVCv070+51-150 - Combined]{\includegraphics[width=.49\textwidth]{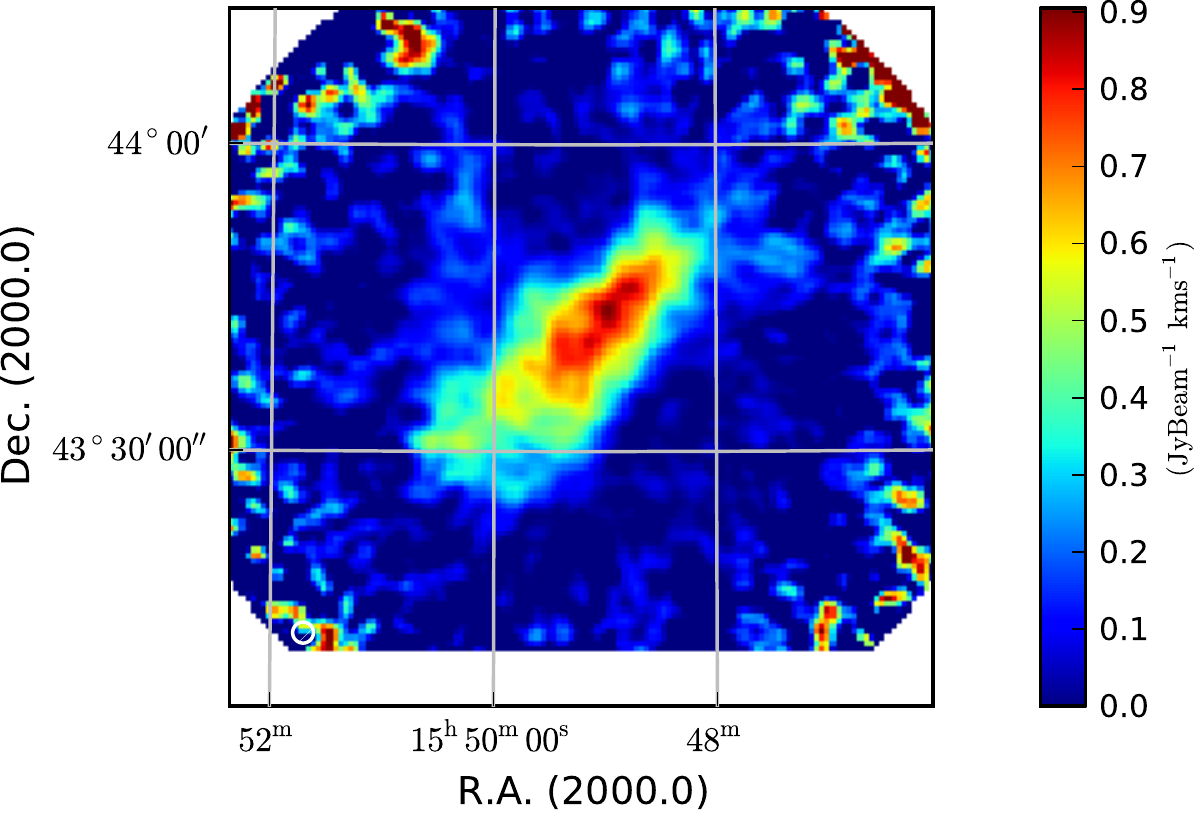}}
\subfigure[Velocity distribution of CHVC\,70+51-150]{\includegraphics[width=0.49\textwidth]{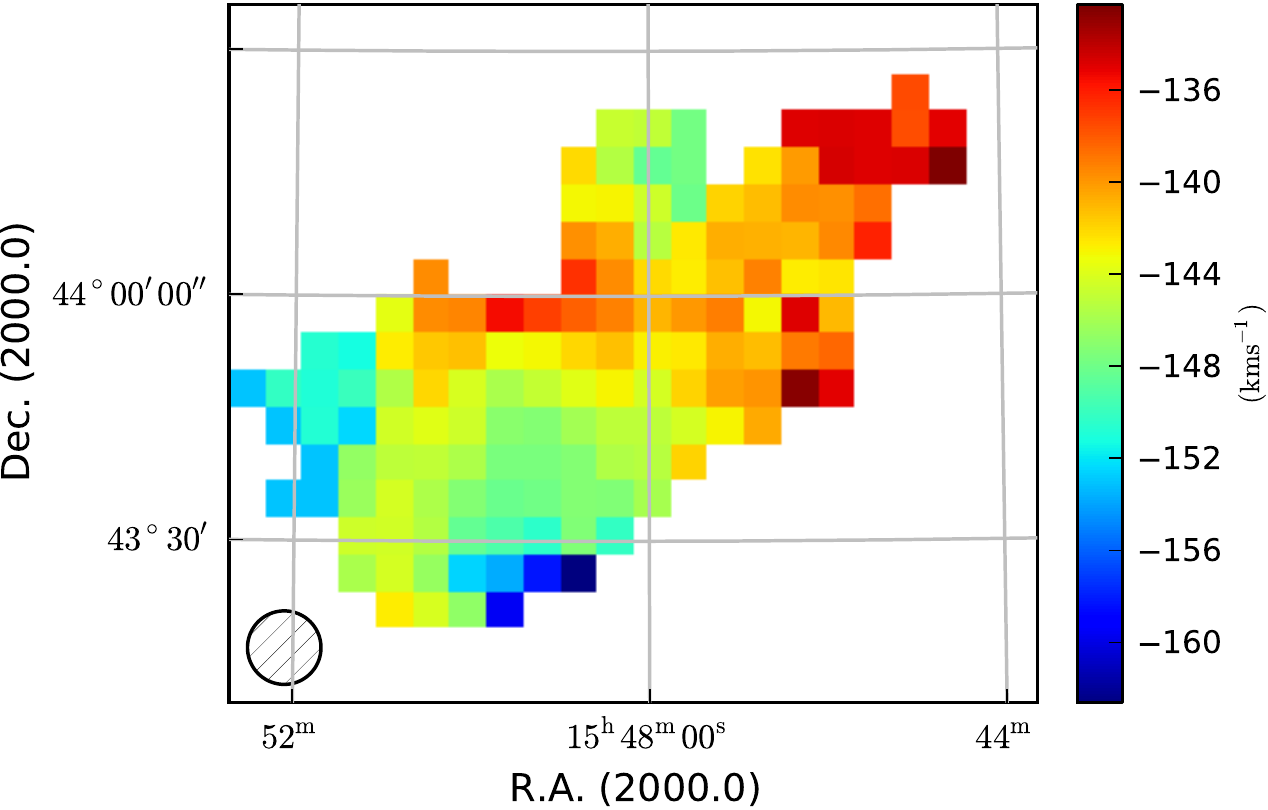}}
\caption{Maps of CHVC\,070+51-150: (a) regridded flux density map of the Effelsberg data, (b) flux-density map of the WSRT data, (c) the combined flux density map, and (d) velocity distribution of CHVC\,70+51-150 in the Effelsberg data cube. The measured velocity gradient in the Effelsberg data cube is about 15 $\mathrm{km\,s^{-1}}$ with a spectral resolution of 5.15 $\mathrm{km\,s^{-1}}$.} 
\label{fig:chvc-070+51}
\end{figure*}

Figure\,\ref{fig:chvc-070+51} comprises different representations of CHVC\,070+51-150 data. Panel (a) shows the regridded Effelsberg \hi 21-cm flux-density map, panel (b) the corresponding WSRT map, panel (c) the flux-density map of the combination, and panel (d) the color-coded first-moment map. The presented weighted radial velocity distribution allows the study of overall velocity distribution across the cloud. For CHVC\,070+51-150 the determined velocity gradient is about 15\,$\mathrm{km\,s^{-1}}$ with a spectral resolution of 5.15\,$\mathrm{km\,s^{-1}}$ (see Table\,\ref{tb:obs_par})

The WSRT data offers a synthesized beam of $\approx 2'\times 2'$. 
The combined data cube has the same high angular and spectral resolution as the interferometric data set. The rms\footnote{The rms values are measured in the data cubes, which are not primary beam corrected.} in the WSRT cube is $1.4\, \mathrm{mJy\,beam^{-1}}$, which is slightly lower than in the combined data ($1.6\,\mathrm{mJy\,beam^{-1}}$) (see also Table\,\ref{tb:phy_par}).

CHVC\,070+51-150 exhibits a distinct HT structure. A remarkable feature of this cloud is the enveloping faint \ion{H}{i} emission, obvious in particular along the major axis of the cloud, which can be seen clearly in the single-dish image (Fig.\,\ref{fig:LR}, top panel).

The measured velocity gradient of about 15\,$\mathrm{km\,s^{-1}}$ (the tail region is slightly slower) present in the first-moment map supports an interaction scenario. The small structures around the cloud and particularly in the northwest region of the cloud have radial velocities slightly slower than the cloud bulk motion, suggesting the cloud structure results from ram pressure interaction.

Table\,\ref{tb:obs_par} summarizes the observational results for CHVC\,070+51-150, while Table\,\ref{tb:phy_par} lists the physical properties of the cloud. In the case of primary beam corrected interferometer data, we have a limited field of view which does not cover the whole cloud. This is necessarily also the case for the combined data. For a strictly fair comparison of the interferometer with the single dish, we consider the region of overlap, since the single-dish map is considerably larger than the interferometric map. The flux determination and comparison should also be handled with care. In order not to choose any artificial cutoffs, we measure the flux from the center of the pointings in ever larger radii (see Fig. \ref{fig:radialprof}). The dashed line marks the HPBW of the WSRT. The results reveal that in the outer regions dominated by diffuse gas, the interferometer misses flux, and its curve flattens out before it is dominated by noise. This shows that for the presented objects the interferometer observations hardly measure flux in the regions dominated by low temperature emission. 

Figure \,\ref{fig:radialprof} (top left) shows the result of CHVC\,070+51-150, where the blue line gives the values for the combination, the green line represents the WSRT values, and the red line the Effelsberg values. The dashed vertical line at 18.5 arcmin marks the boundary of the primary beam of WSRT (effective beam size is $\approx\,37'$). 

The course of the curves can be explained as follows: the high density clumps are concentrated mainly in the central part of the cloud. Accordingly, the radio interferometric map exhibits higher brightness temperatures than the single-dish. For larger radial distances from the pointing center there is more diffuse gas and therefore the values in the single-dish and combined cubes are higher. The total flux in the combined map is quantitatively consistent with the total flux measured with the Effelsberg telescope. The dashed line marks the HPBW of the WSRT. About 20\% more flux has been detected by the single-dish telescope than by the radio interferometer. This result indicates that a significant fraction of the cloud consists of diffuse extended \ion{H}{i} missed by the interferometer as a result of its insensitivity at the lowest spatial frequencies.\\

\begin{table*}[ht]
\centering
\begin{tabular}{l c c c c c c c}
\toprule
Name          & Peak $T_\mathrm{B}$	  & Peak $N_\mathrm{\ion{H}{i}}$    &  minor x major ext. &   WNM \\
              &                         &                                 & comb.               & Eff/WSRT/comb. \\
	          & [K]                     & $[10^{19}\,\mathrm{cm^{-2}]}$   &  [$^\circ$]         & [$\mathrm{km\,s^{-1}}$] \\
\midrule
CHVC\,070+51-150  & 1.1                     & 4.5                             & 0.2 x 3.3           & 22.8/19.2/19.6 \\
CHVC\,108-21-390  & 0.7                     & 3.2                             & 0.3 x 1.0           & 21.4/17.4/17.2 \\
CHVC\,164+03-186  & 1.4                     & 5.5                             & 0.6 x 1.6           & 22.5/19.2/18.6 \\
\bottomrule
\end{tabular}
\caption{Physical properties of the CHVCs. Peak $T_B$ is the peak brightness temperature; peak $N_{\rm HI}$ denotes the maximum column density; major and minor  are the angular extent of the clouds; and WNM is the FWHM of the gaseous components in the combined data. The quantities are given separately for the Effelsberg, WSRT, and combined data.}
\label{tb:phy_par}
\end{table*}

\subsection{CHVC\,108-21-390} \label{subsec:108-21}
\begin{figure*}[ht]
\centering
\subfigure[CHVC\,108-21-390 $\mathrm{Effelsberg_{\,reg}}$]{\includegraphics[width=0.48\textwidth]{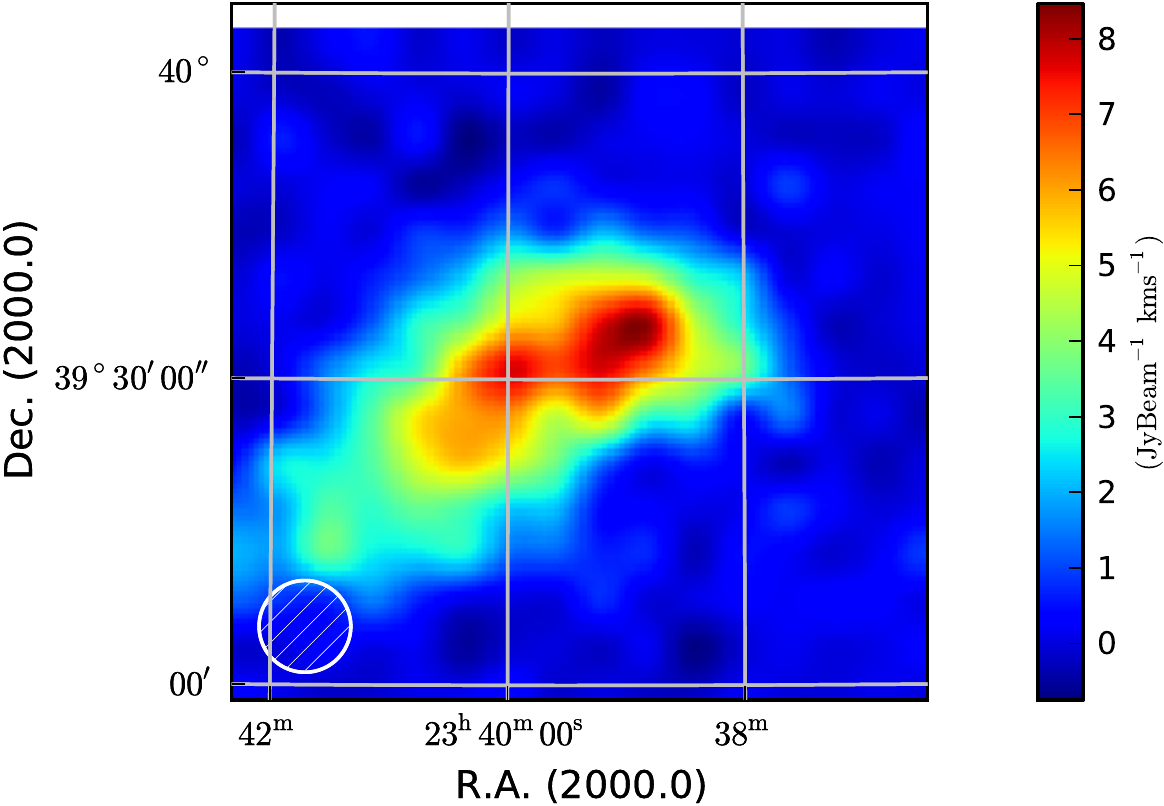}}
\subfigure[CHVC\,108-21-390 WSRT]{\includegraphics[width=0.49\textwidth]{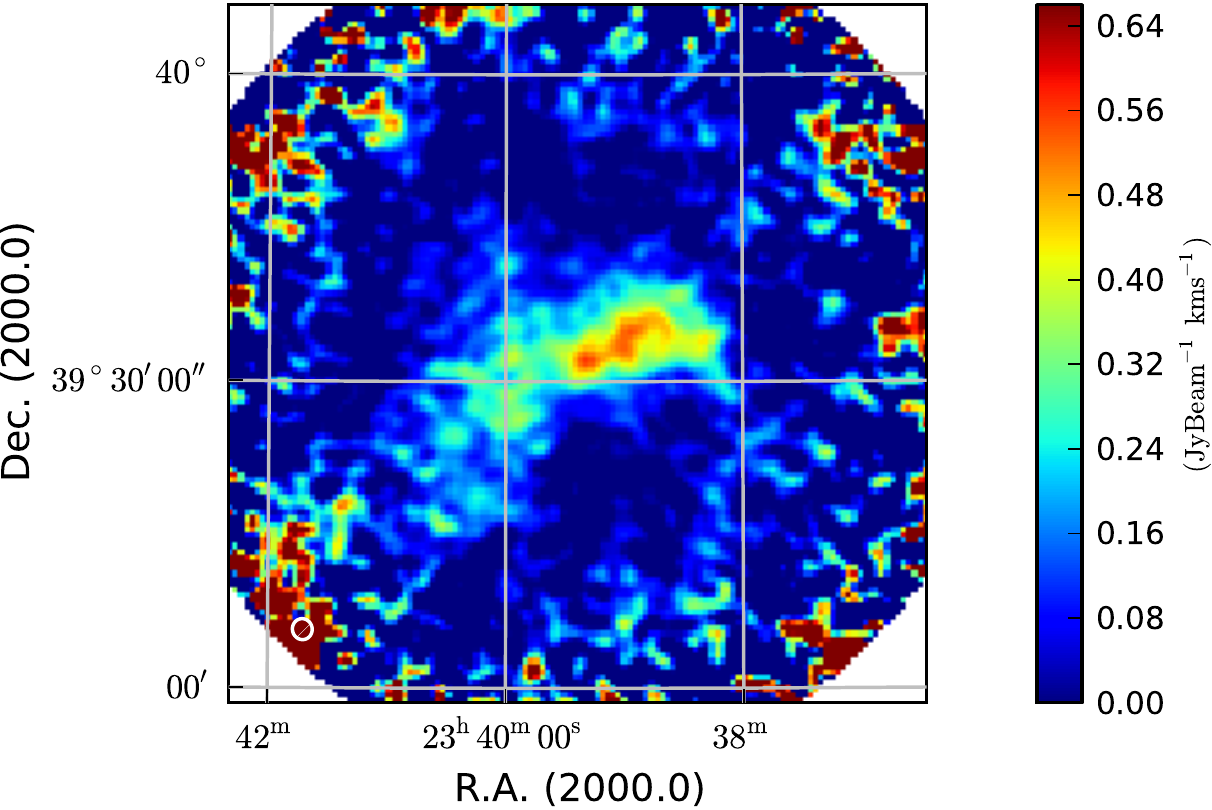}}
\subfigure[CHVC\,108-21-390 Combined]{\includegraphics[width=0.49\textwidth]{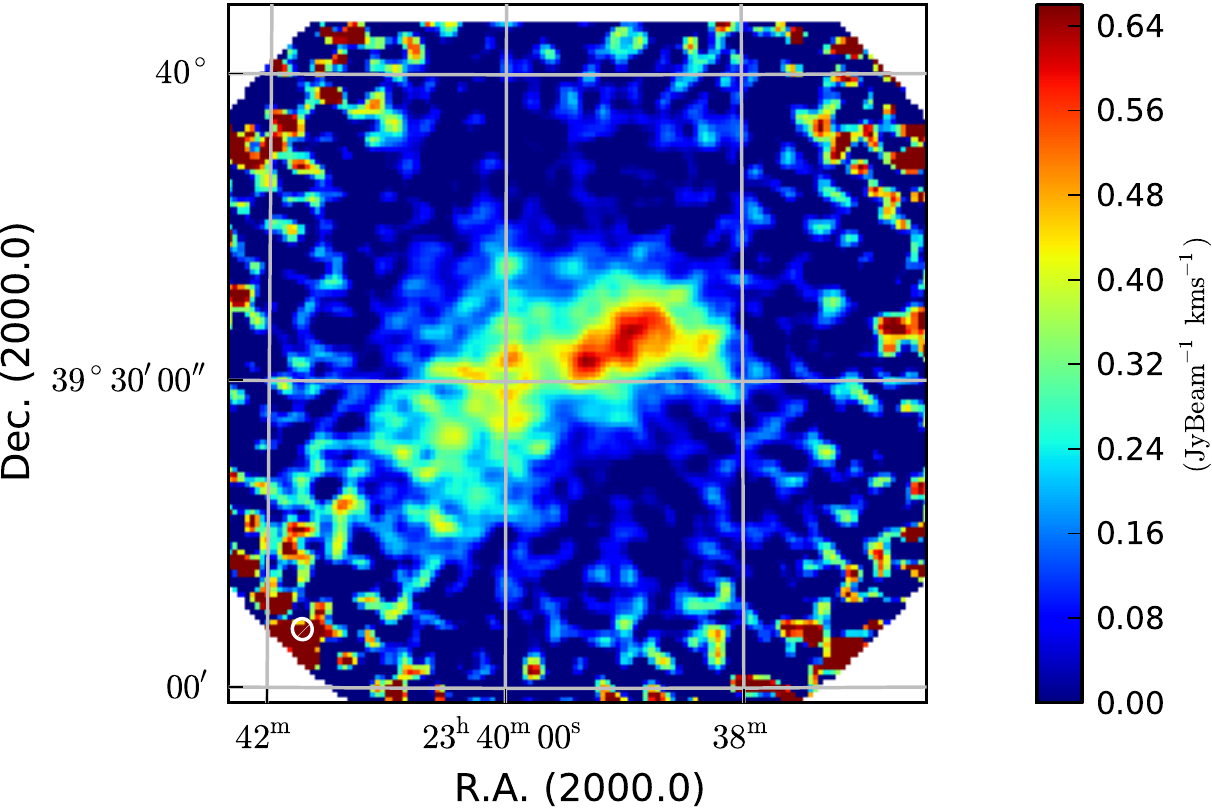}}
\subfigure[Velocity distribution of CHVC\,108-21-390]{\includegraphics[width=0.49\textwidth]{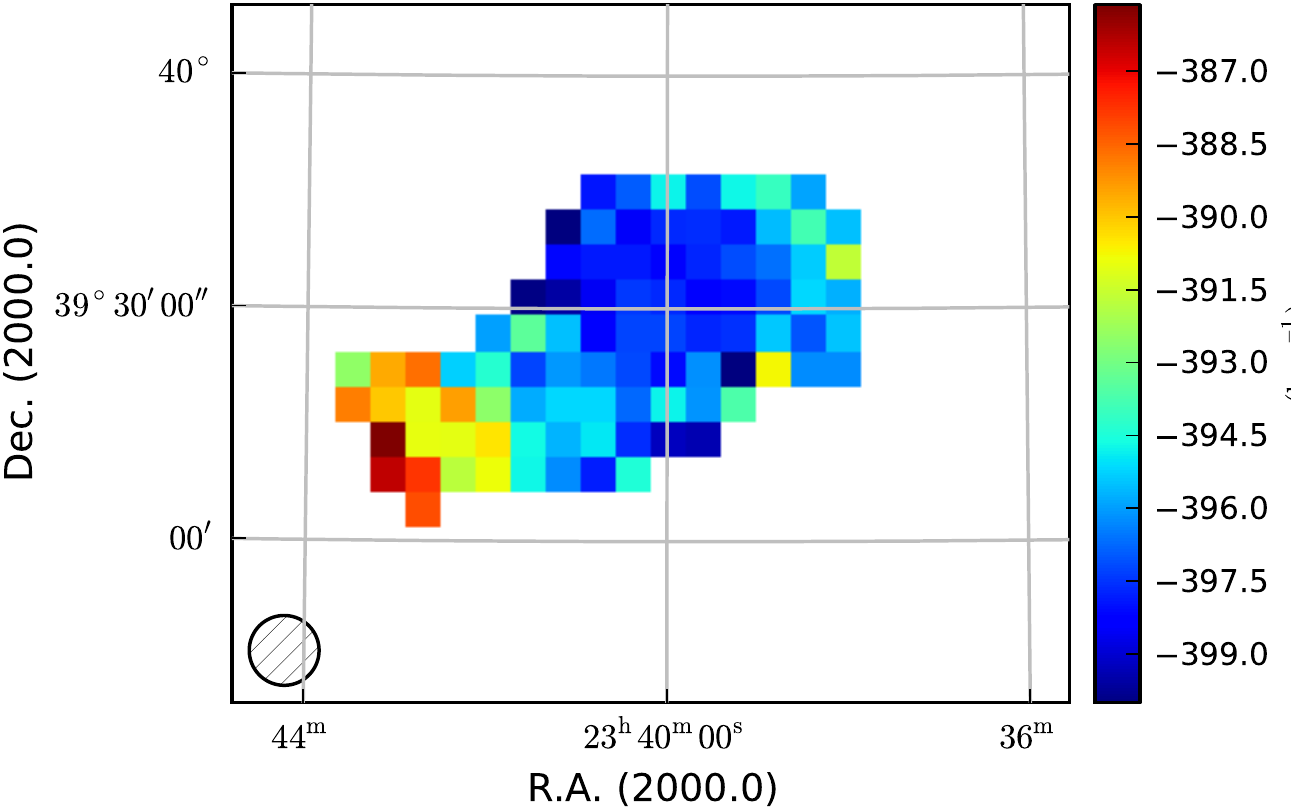}}
\caption{Maps of CHVC\,108-21-390: (a) map of the regridded Effelsberg data, (b) the WSRT data, (c) the combined data, and (d) the velocity distribution of CHVC-108-21. The measured velocity gradient within the cloud is $\approx \, 10 \mathrm{km\,s^{-1}}$ with a spectral channel width of 2.57 $\mathrm{km\,s^{-1}}$.}
\label{fig:chvc-108-21}
\end{figure*}

Figure\,\ref{fig:chvc-108-21} presents the corresponding flux density and first-moment maps of CHVC\,108-21-390. Table\,\ref{tb:obs_par} summarizes the basic parameters of the data set while Table\,\ref{tb:phy_par} compiles the derived cloud properties.
The WSRT synthesized beam is about $2.9'$x$1.9'$ in size. This is also the angular resolution of the combined \hi data.

The velocity distribution of CHVC\,108-21-390 indicates bulk velocities of about -400 $\mathrm{km\,s^{-1}}$, which approaches the upper limits of the velocities measured for HVCs \citep{2004ASSL..312...25W, chvc2004}. 
The head of the cloud is located at the northwest and the lower flux density tail is located at the southeast (Fig.\,\ref{fig:chvc-108-21}, panels (a) and (c)). The faint diffuse gas at the southeast edge is visible indistinctly in the interferometer map (Fig.\,\ref{fig:chvc-108-21}, top panel (b)), yet it can be seen clearly in both single-dish (panel (a)) and combined data (panel (c)). The angular major axis of the cloud is approximately $1^{\circ}$, while the minor axis is about $0.3^{\circ}$ (see Table\,\ref{tb:phy_par}). 
The velocity distribution indicates a velocity gradient of about 10 $\mathrm{km\,s^{-1}}$ with the slower gas located in the tail. After the combination we measure $\approx\,40\%$ difference with respect to the total fluxes in the interferometric and combined data (Fig.\,\ref{fig:radialprof}, middle panel). The difference of the total flux in single-dish and radio interferometer reveals that almost half the cloud consists of diffuse extended \hi emission, which is missed by the WSRT. 

\subsection{CHVC\,162+03-186} \label{subsec:162+03}
\begin{figure*}[ht]
\centering
\subfigure[CHVC\,162+03-186 $\mathrm{Effelsberg_{\,reg}}$]{\includegraphics[width=0.48\textwidth]{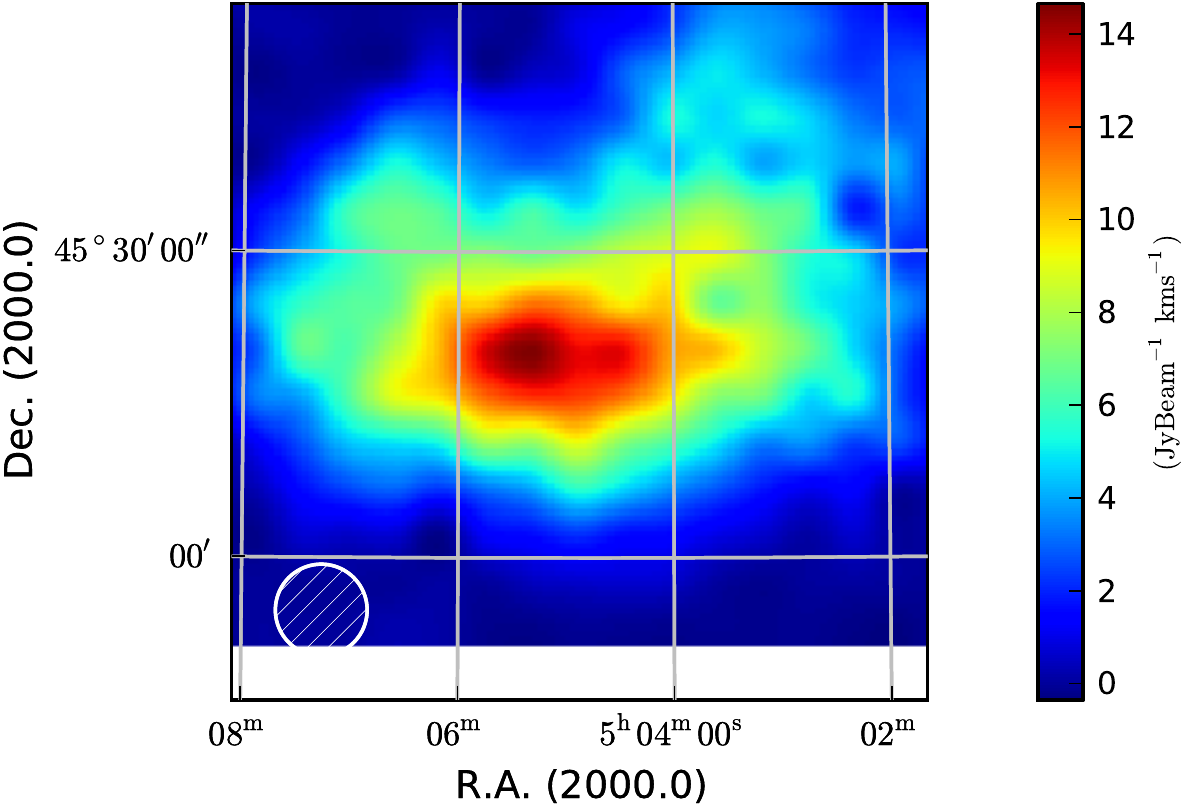}}
  \subfigure[CHVC\,162+03-186 interferometer]{\includegraphics[width=0.49\textwidth]{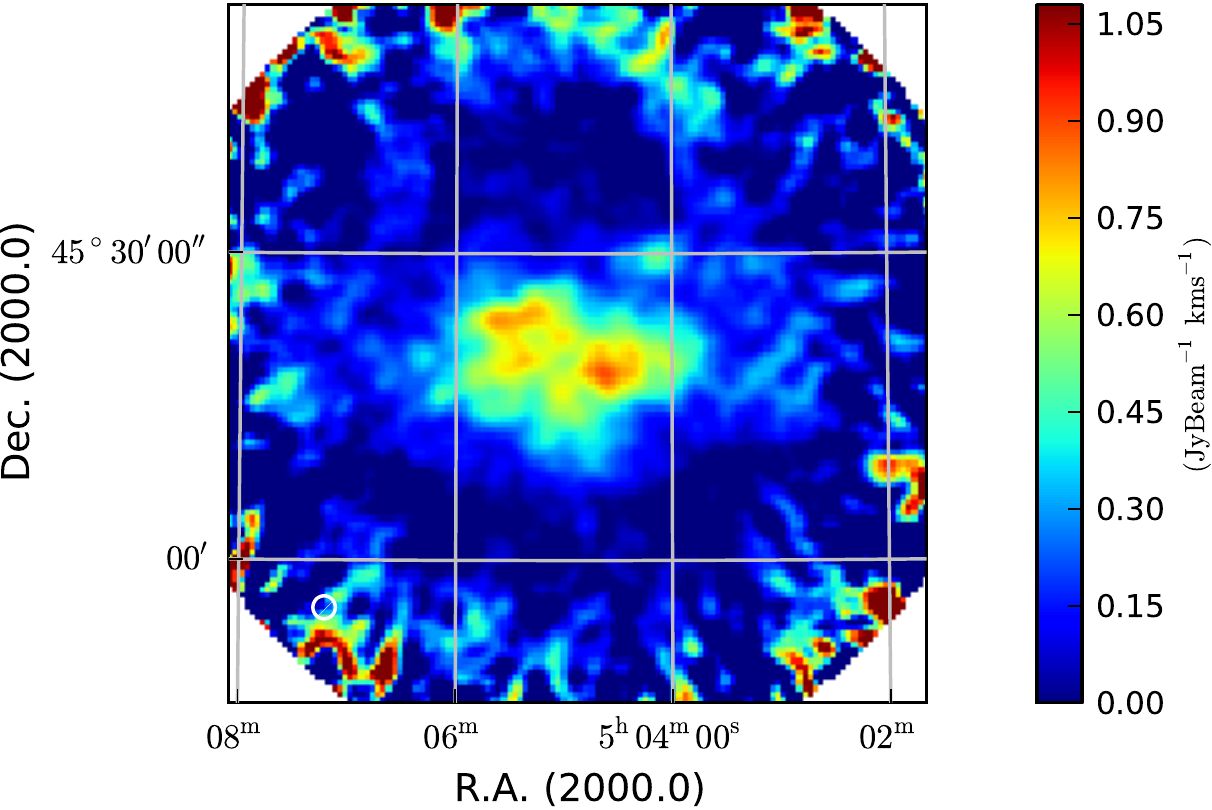}}
  \subfigure[CHVC\,162+03-186 combined]{\includegraphics[width=0.49\textwidth]{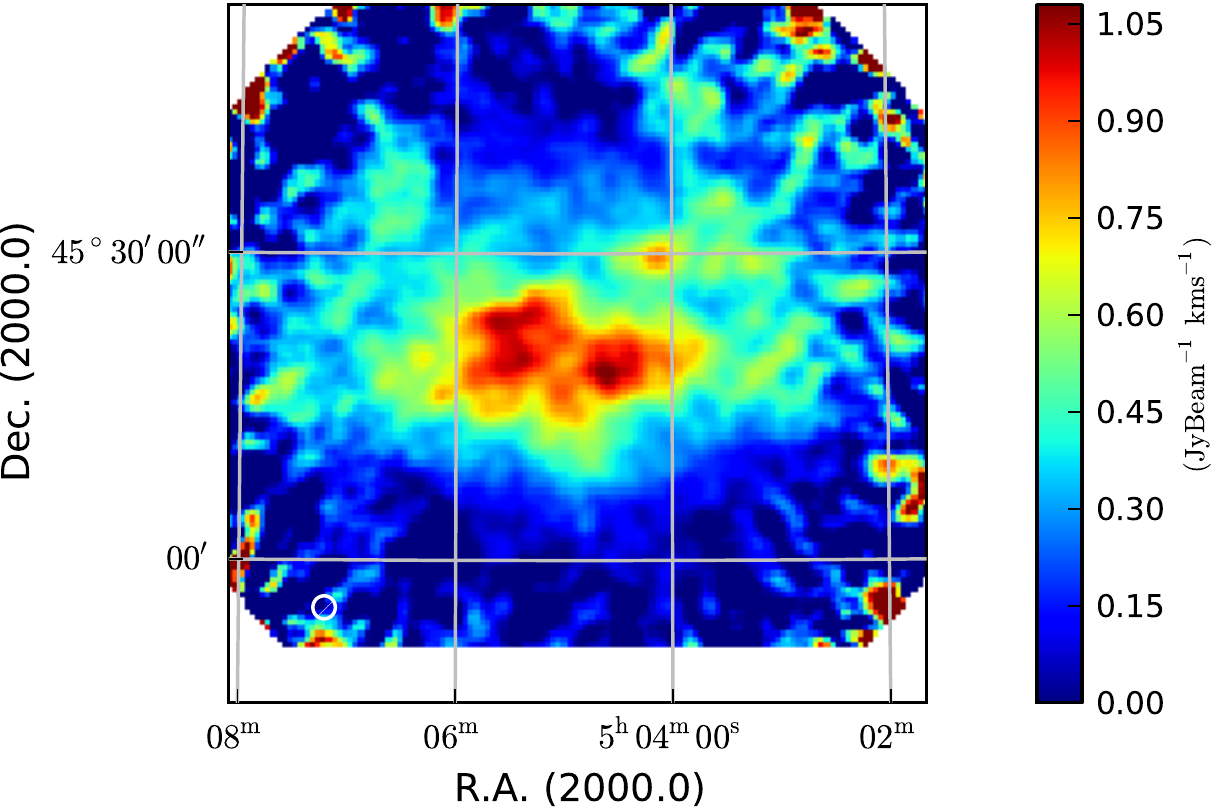}}
  \subfigure[Velocity distribution of CHVC\,162+03-186]{\includegraphics[width=0.49\textwidth]{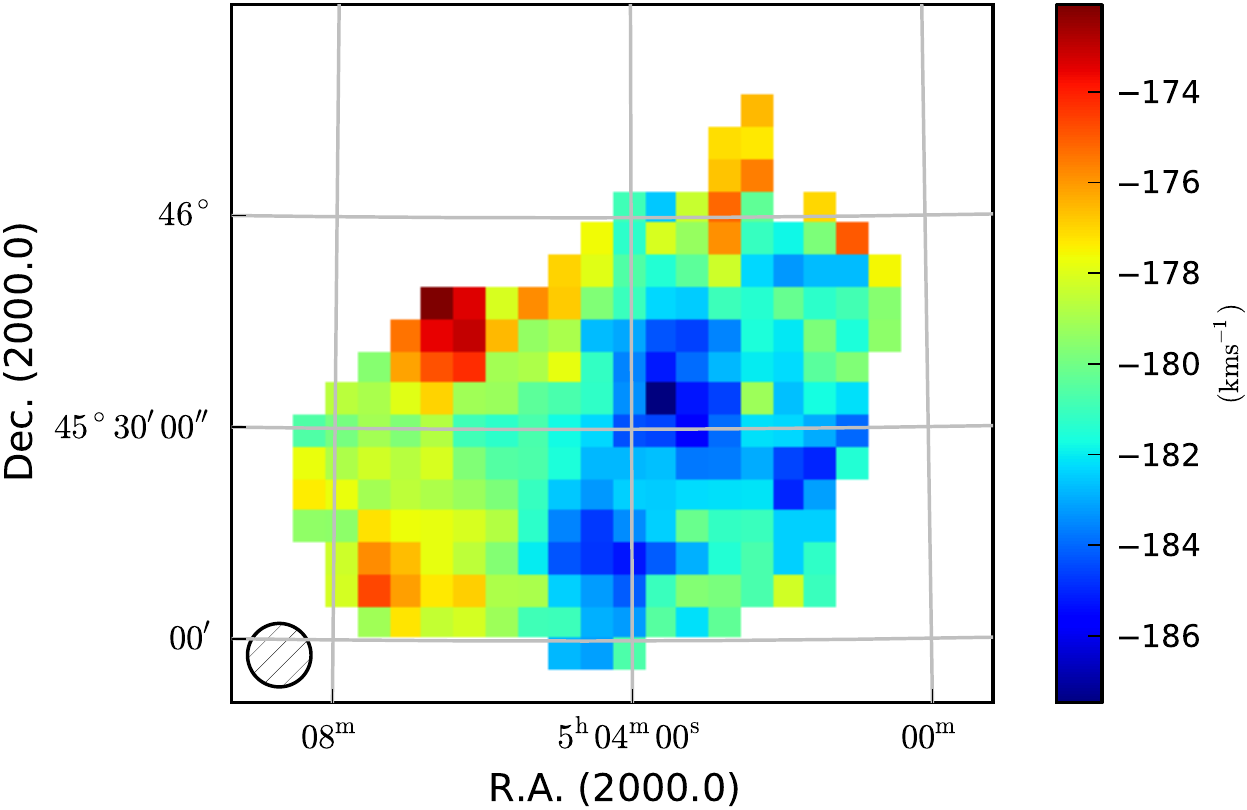}}
  \caption{Maps of CHVC\,162+03-186: (a) the regridded Effelsberg data, (b) flux-density map of the WSRT data, (c) the combined flux density map, and (d) the flux weighted velocity distribution with a gradient of 15 $\mathrm{km\,s^{-1}}$ and spectral channel width of 2.57 $\mathrm{km\,s^{-1}}$.} 
\label{fig:chvc-162+03}
\end{figure*}

Figure\,\ref{fig:chvc-162+03} presents the \hi maps of CHVC\,162+03-186 and is arranged as the previous figures are. Table\,\ref{tb:obs_par} compiles the observational parameters while Table\,\ref{tb:phy_par} summarizes the derived physical properties. \citet{2005A&A...432..937W} found that CHVC\,162+03-186 has a bow-shock shape (Fig. \ref{fig:LR}). The bow-shock might be the result of interactions between the cloud and the gaseous halo. For the WSRT data the synthesized beam is $2.2'$ x $2.1'$. The combined map has the same angular and spectral resolution as the WSRT data.
Figure\,\ref{fig:chvc-162+03} panel (d) displays the velocity distribution of CHVC\,162+03-186. It shows velocities of $\approx -180 \,\mathrm{km\,s^{-1}}$. The slower gas is located mostly at the eastern edge of the cloud. There are two distinct regions, separated spatially, that have higher velocities of about $-190\,\mathrm{km\,s^{-1}}$.

\begin{figure*}[htbp]
\centering
\includegraphics[width=0.43\textwidth]{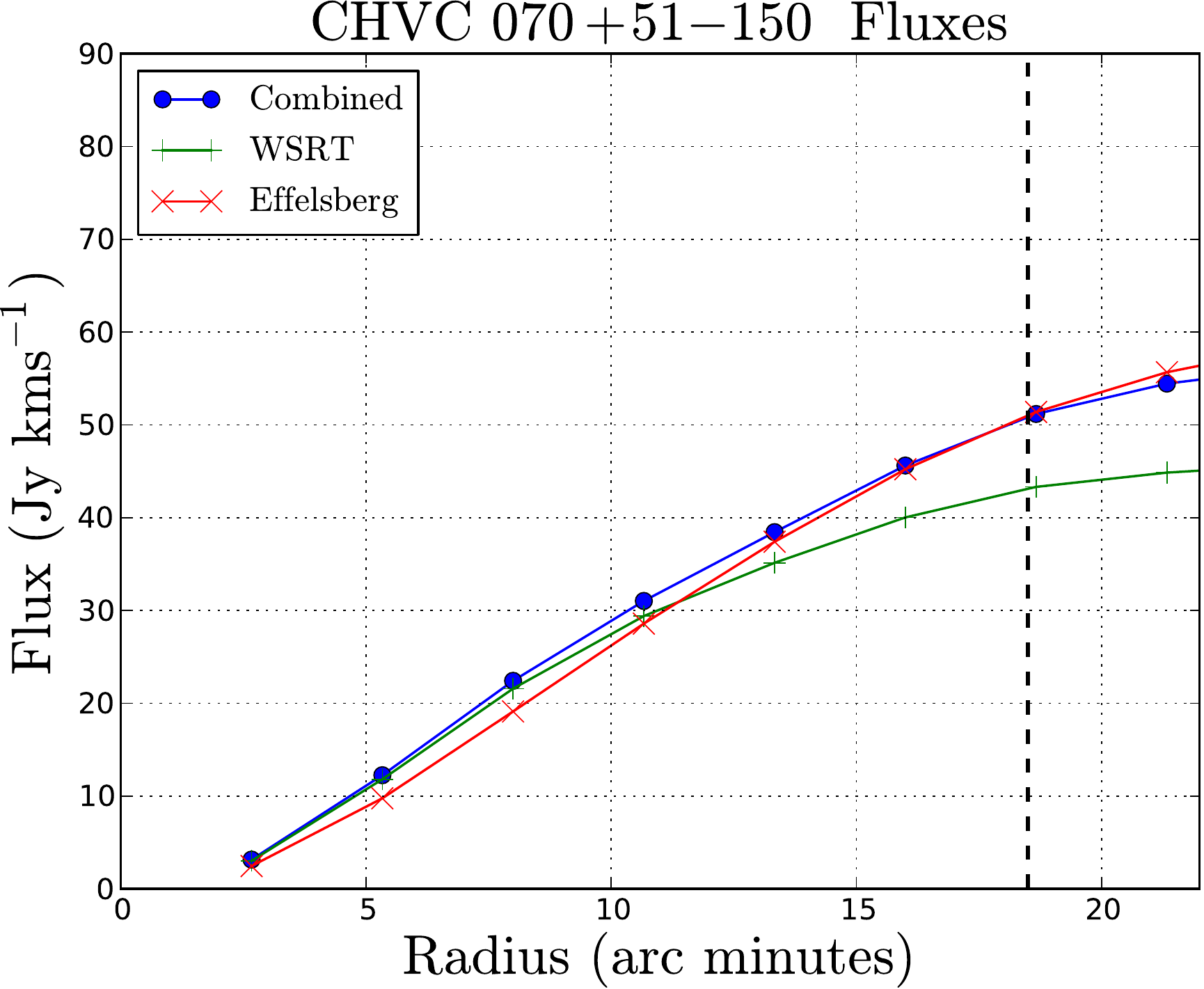}
\includegraphics[width=0.43\textwidth]{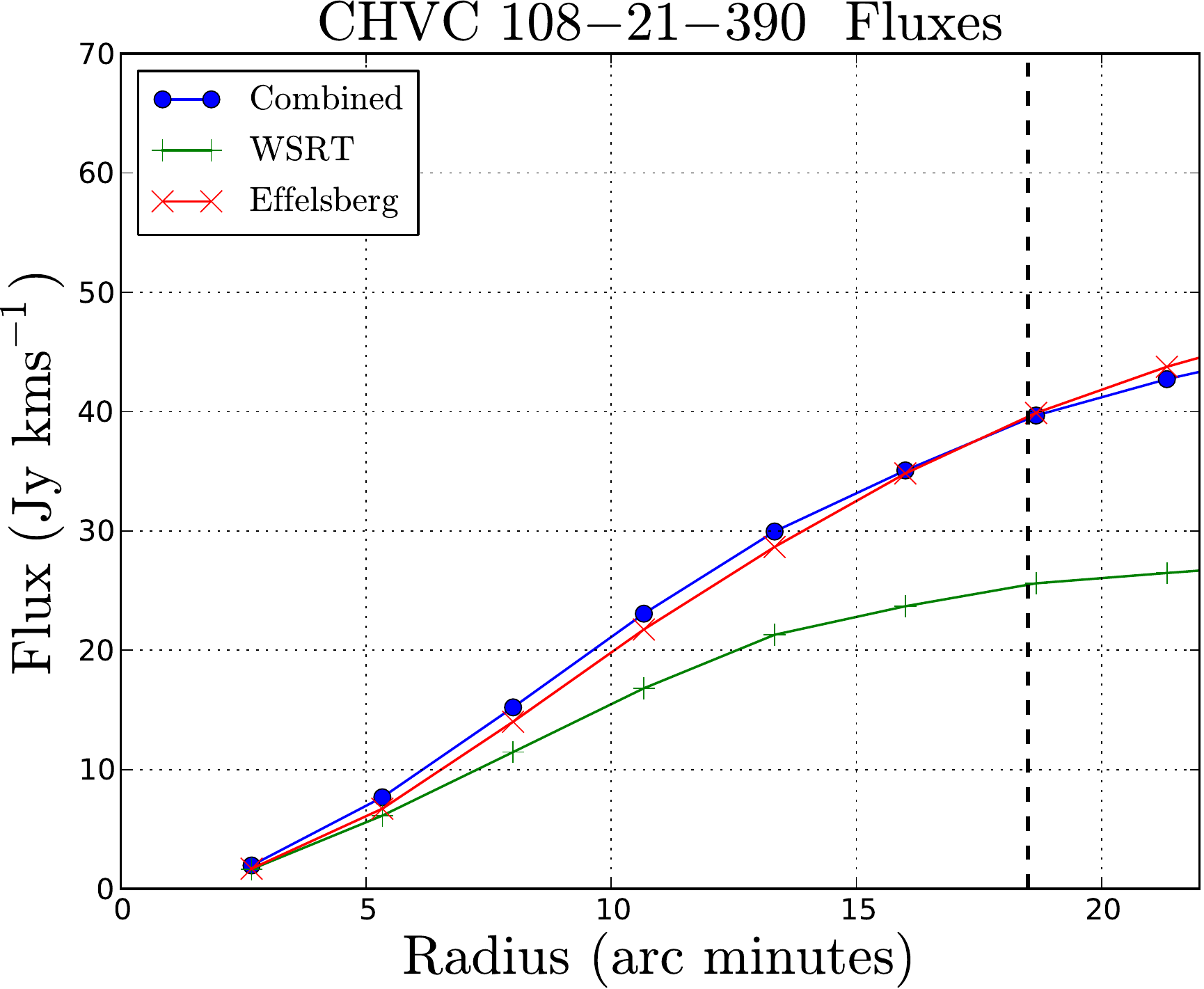}
\includegraphics[width=0.44\textwidth]{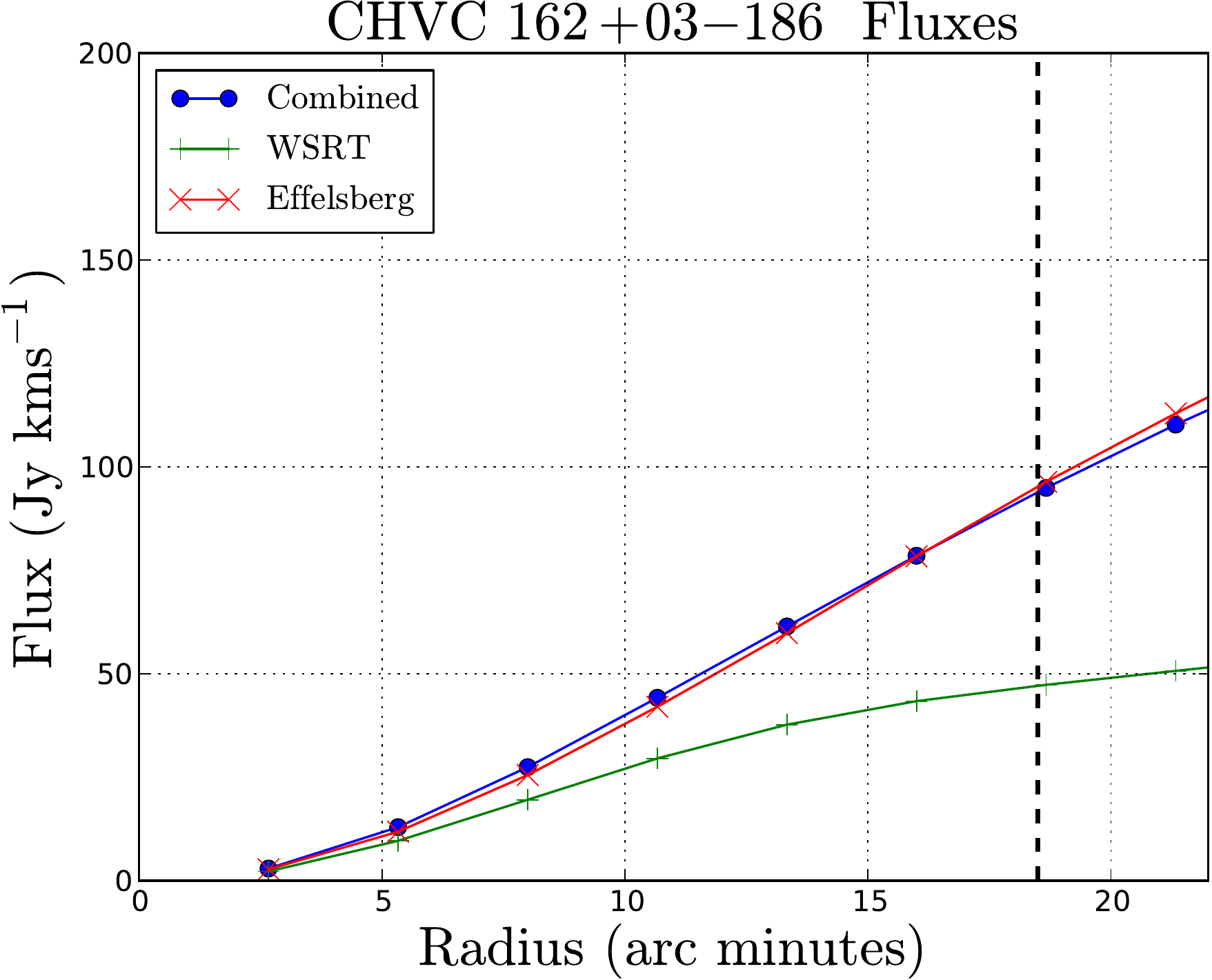}
\caption{Measured cumulative fluxes as a function of radial separation from the centre of the maps. The top-left panel presents the results of radial profiles for CHVC\,070+51-150, the top-right panel for CHVC\,108-21-390, and the bottom panel for CHVC\,162+03-186. The blue line represents the measured fluxes for the combination, the green line the values for WSRT, and the red line Effelsberg. The dashed line marks the HPBW of the WSRT. In all three cases the measured flux after the combination is in good agreement with the measured value from the single-dish data. In the case of CHVC\, 070+51, about $20\%$ more flux has been detected by the single-dish than by the radio interferometer. The difference between the combined and interferometric data is $\approx \, 40\%$ for CHVC\,108-21-390. For CHVC\,162+03-186, the ratio of interferometric/combined accounts for almost $50\%$. This is the largest measured ratio for the three CHVCs. It also reveals that a large portion of the cloud consists of WNM which is traced best by the single-dish. The results demonstrate that the combination achieves the expectations.}
\label{fig:radialprof}
\end{figure*}

Figure\,\ref{fig:radialprof} shows the cumulative fluxes of the three data sets against the concentric rings from the center of the map. Again the dashed line indicates the primary beam of the WSRT. A comparison between measured total flux reveals a $\approx\,50\%$ higher total flux density in the combined data than in the interferometric data. This large difference suggests the existence of a substantial amount of broadly distributed diffuse gas in CHVC\,162+03-186. 

\subsection{Gaussian decomposition} \label{subsec:decomp}

Head-tail clouds often show a two-phase medium \citep{2000A&A...357..120B, 2005ASPC..331..105W, 2006A&A...457..917B}.
In the case of the CHVCs under consideration, the small-scale structure detected by the radio interferometer might also trace denser and probably cooler gas.
A two-component Gaussian decomposition of the \hi data is a useful approach to separate quantitatively these two gas components, where the narrow line width corresponds to the CNM with high volume densities and the broad line width to the low density WNM regions.

To evaluate the relative amount of \hi in the warm and cold gas phase, we create average profiles for each CHVC. Because of the intrinsic velocity gradient, a simple sum over all spectra would yield an artificially broadened \hi line spectrum. To account for this, we use a technique similar to the recently introduced super profiles by \citet{2012AJ....144...96I}. Each line of sight that has at least six channels above the 3-$\sigma$ level is fitted with a Gauss-Hermite polynomial to robustly determine the radial velocity of the peak of the \hi profile. The individual profiles are then shifted to a common radial velocity and summed. The resulting super profile is then fitted with a two-component Gaussian to estimate the relative abundances of the cold and warm phases, i.e. the ratio of the areas of the two Gaussian components.

\begin{figure*}[ht]
\centering
\subfigure[CHVC\,070+51-150 combined]{\includegraphics[scale=0.38]{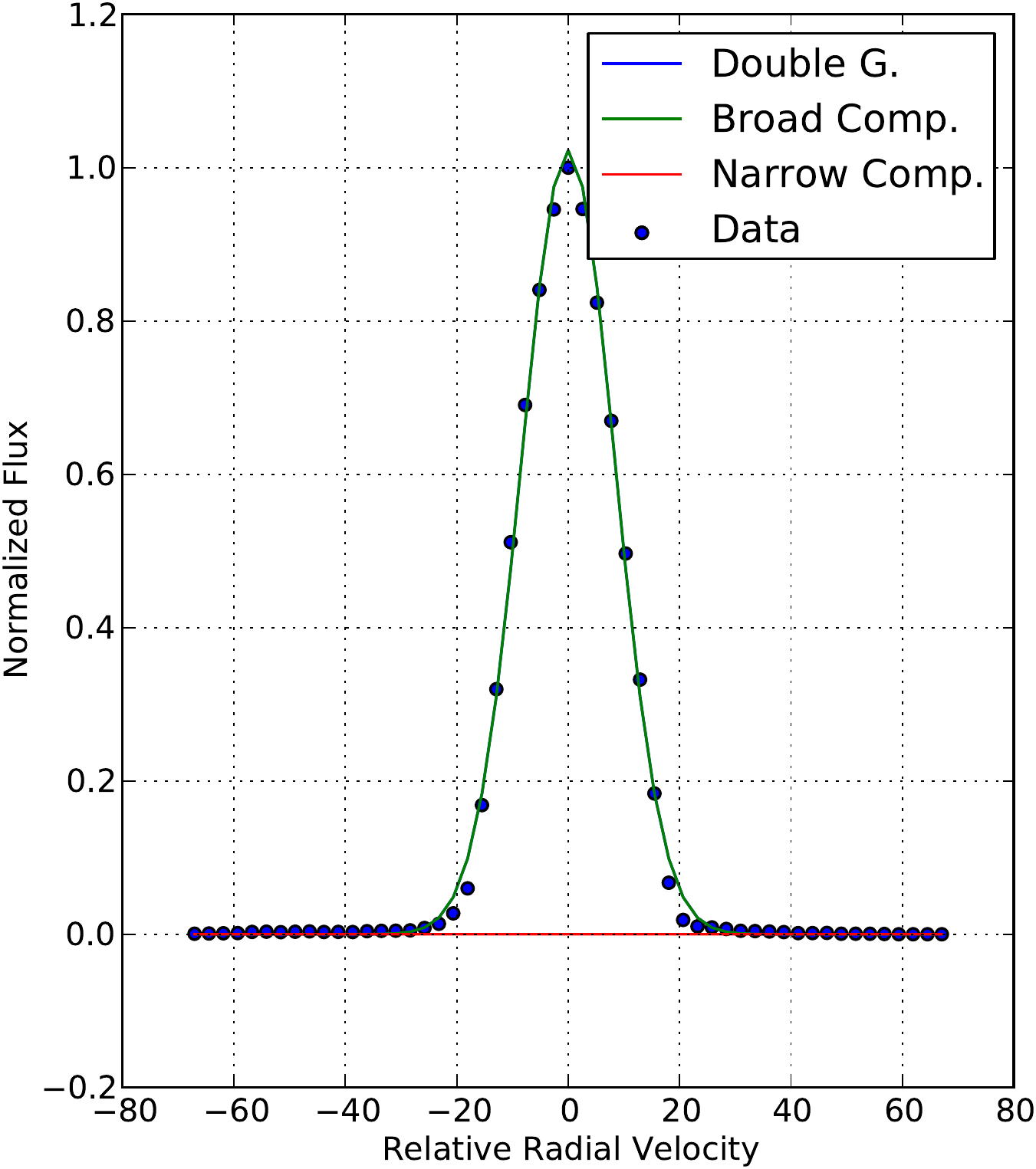}}
\subfigure[CHVC\,070+51-150 WSRT]{\includegraphics[scale=0.38]{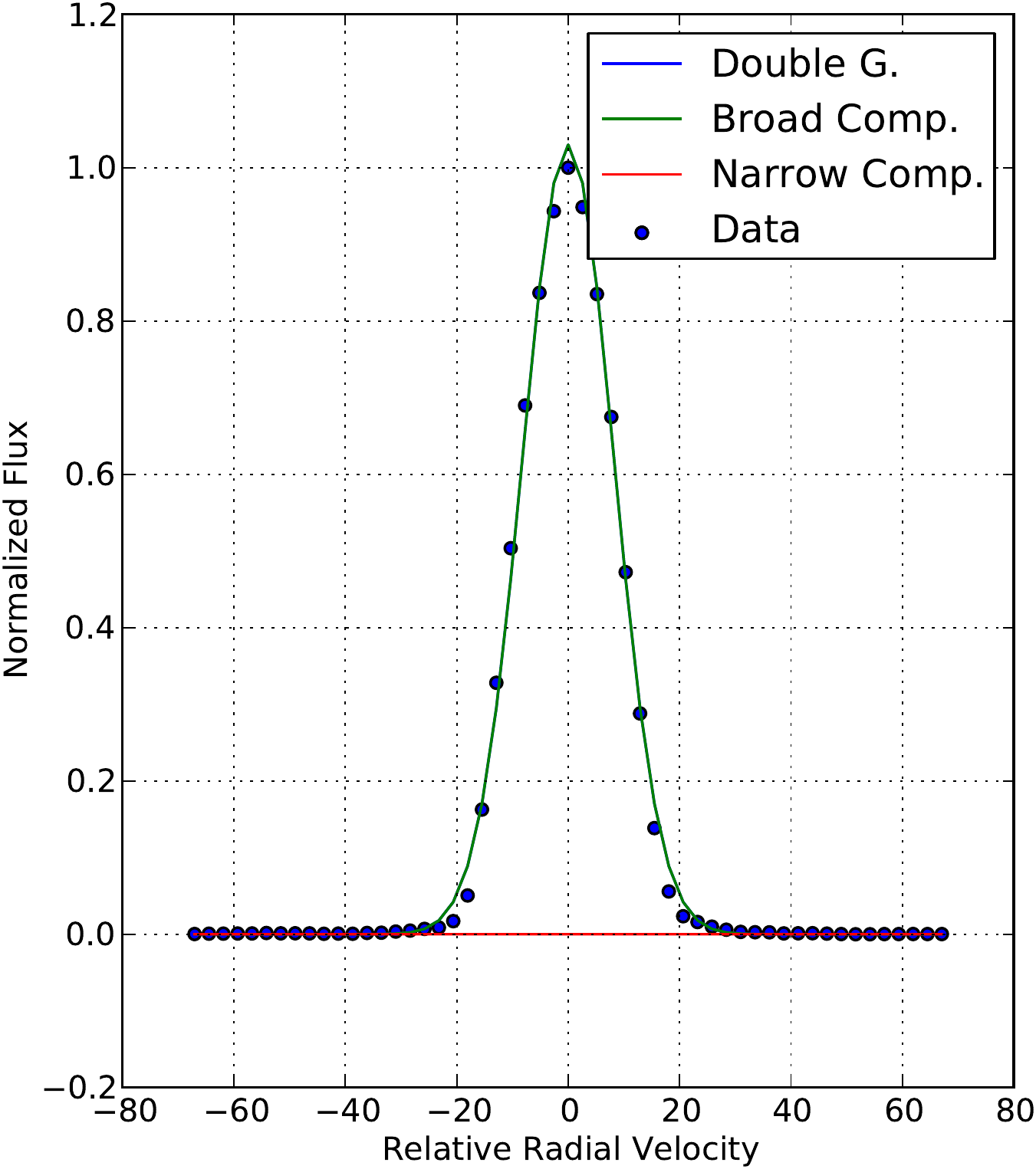}}
\subfigure[CHVC\,070+51-150 $\mathrm{Effelsberg_{\, reg}}$]{\includegraphics[scale=0.38]{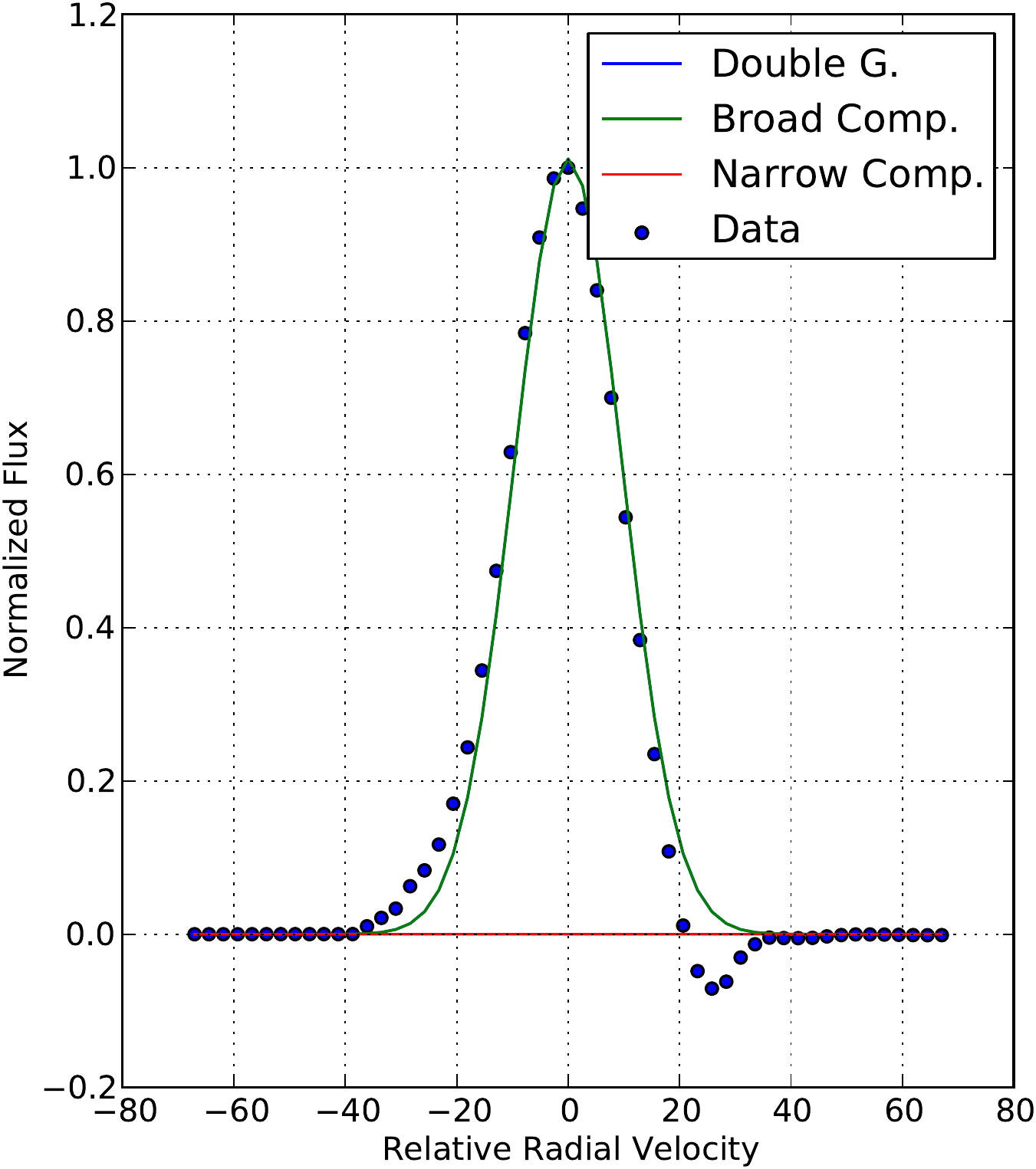}}
%
\subfigure[CHVC\,108-21-390 combined]{\includegraphics[scale=0.38]{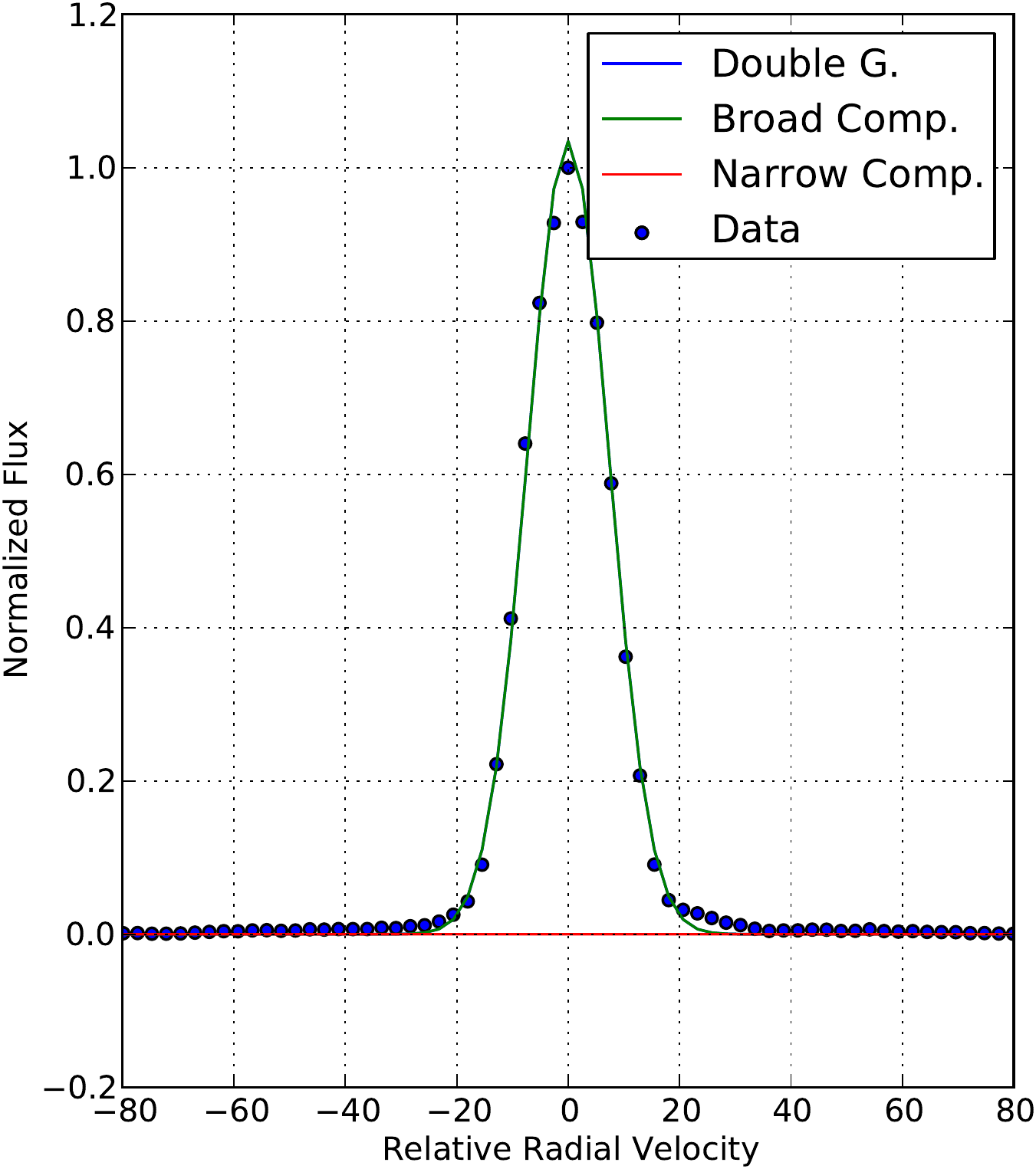}}
\subfigure[CHVC\,108-21-390 WSRT]{\includegraphics[scale=0.38]{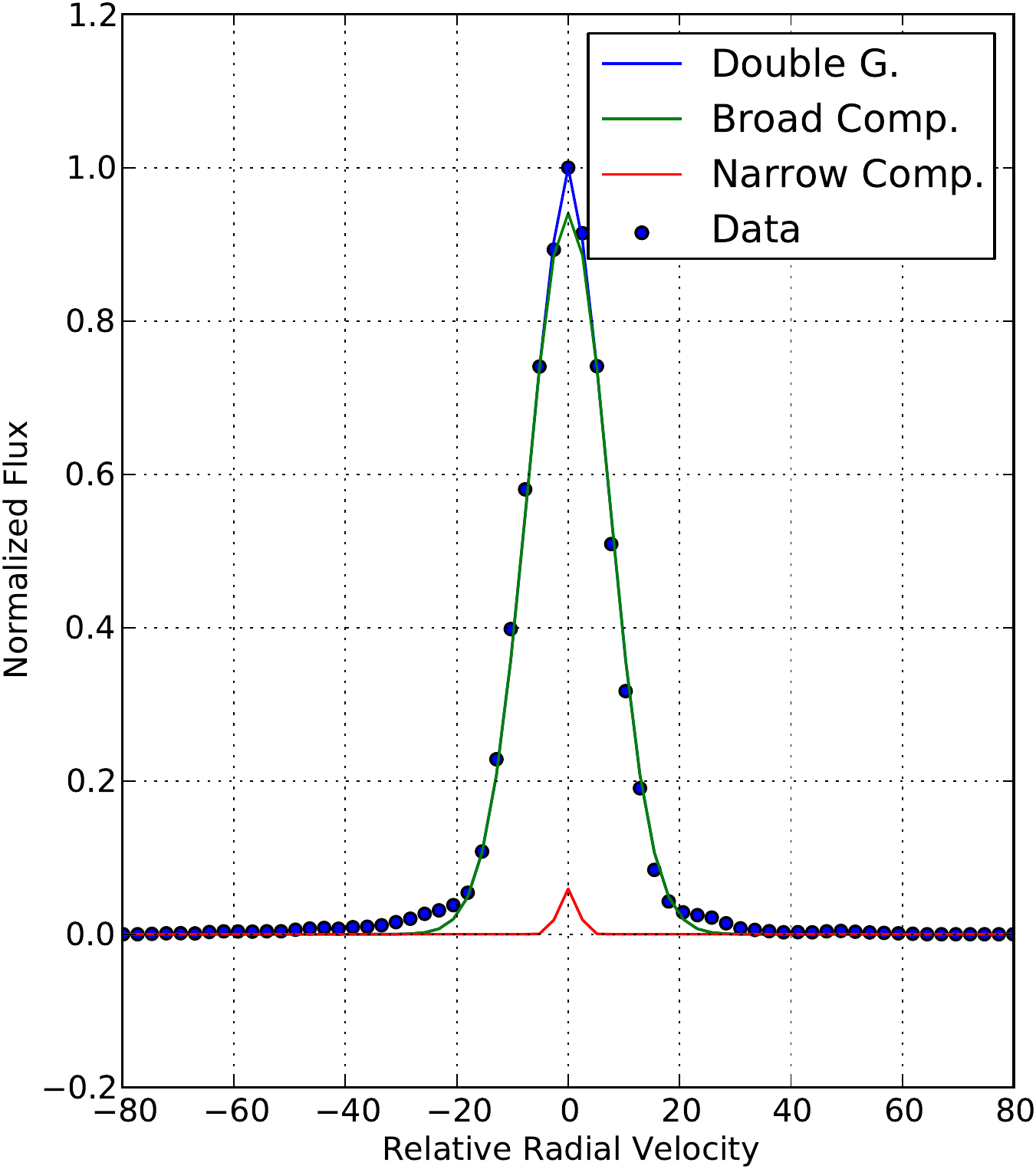}}
\subfigure[CHVC\,108-21-390 $\mathrm{Effelsberg_{\, reg}}$]{\includegraphics[scale=0.38]{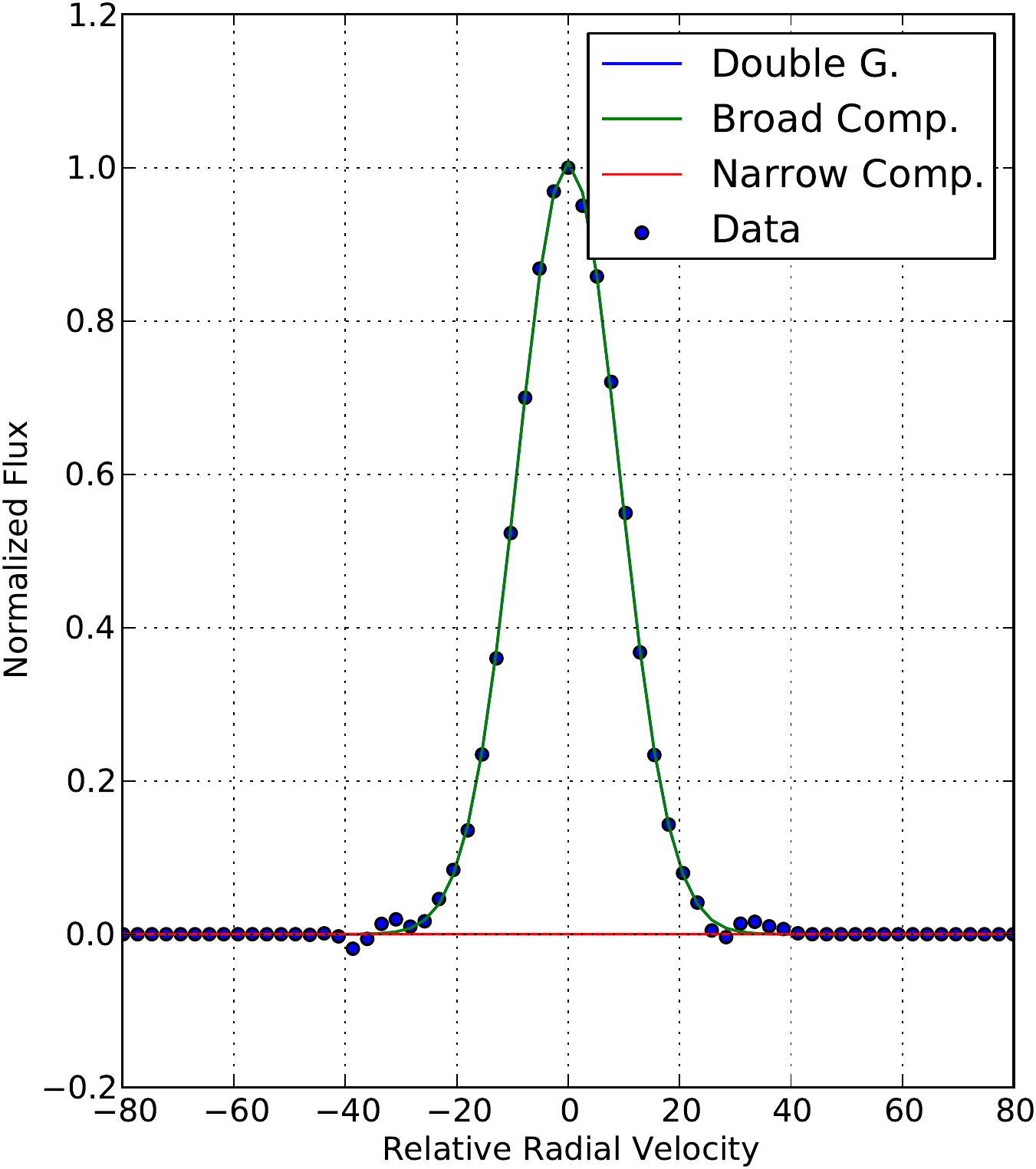}}
%
\subfigure[CHVC\,162+03-186 combined]{\includegraphics[scale=0.38]{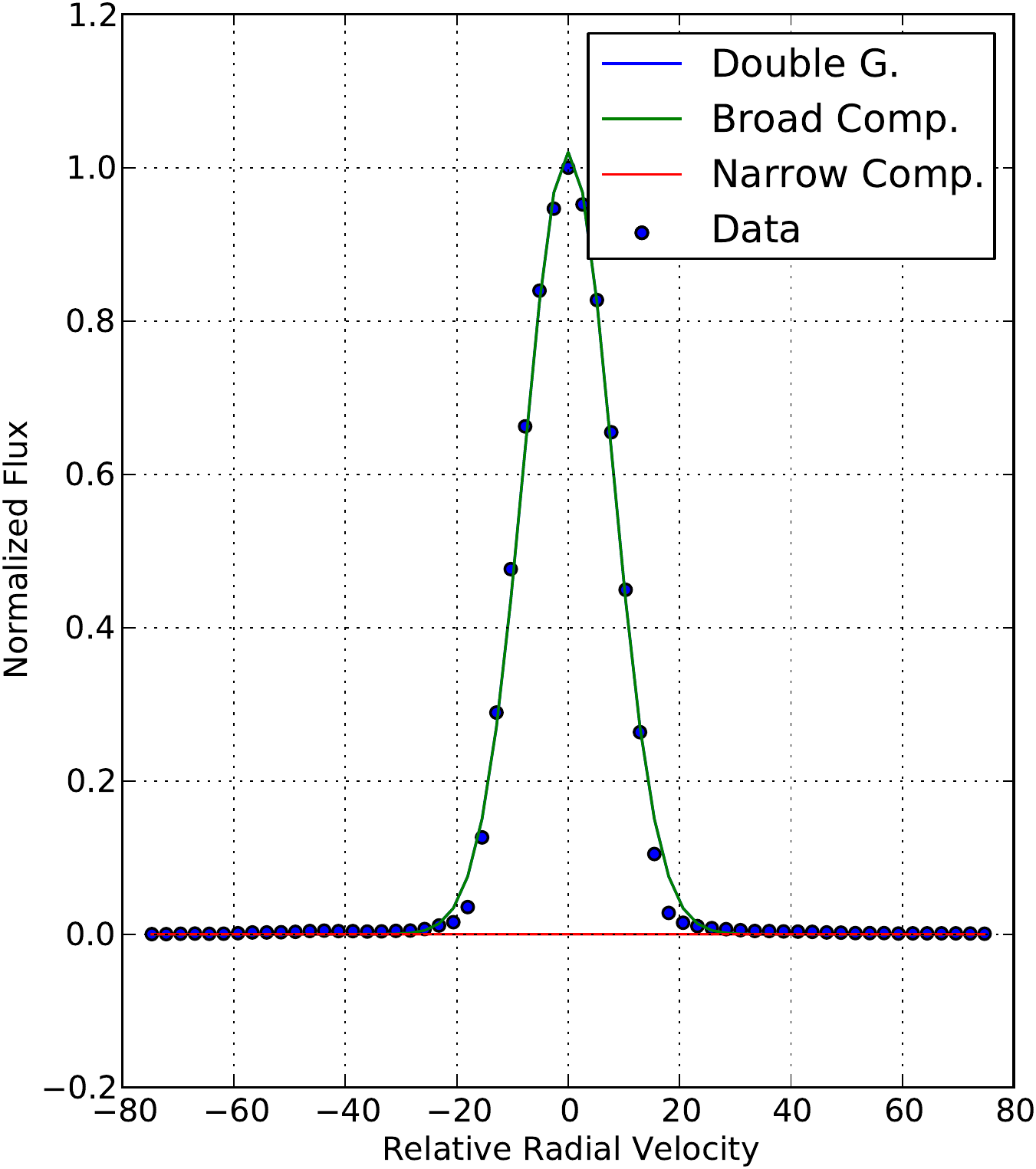}}
\subfigure[CHVC\,162+03-186 WSRT]{\includegraphics[scale=0.38]{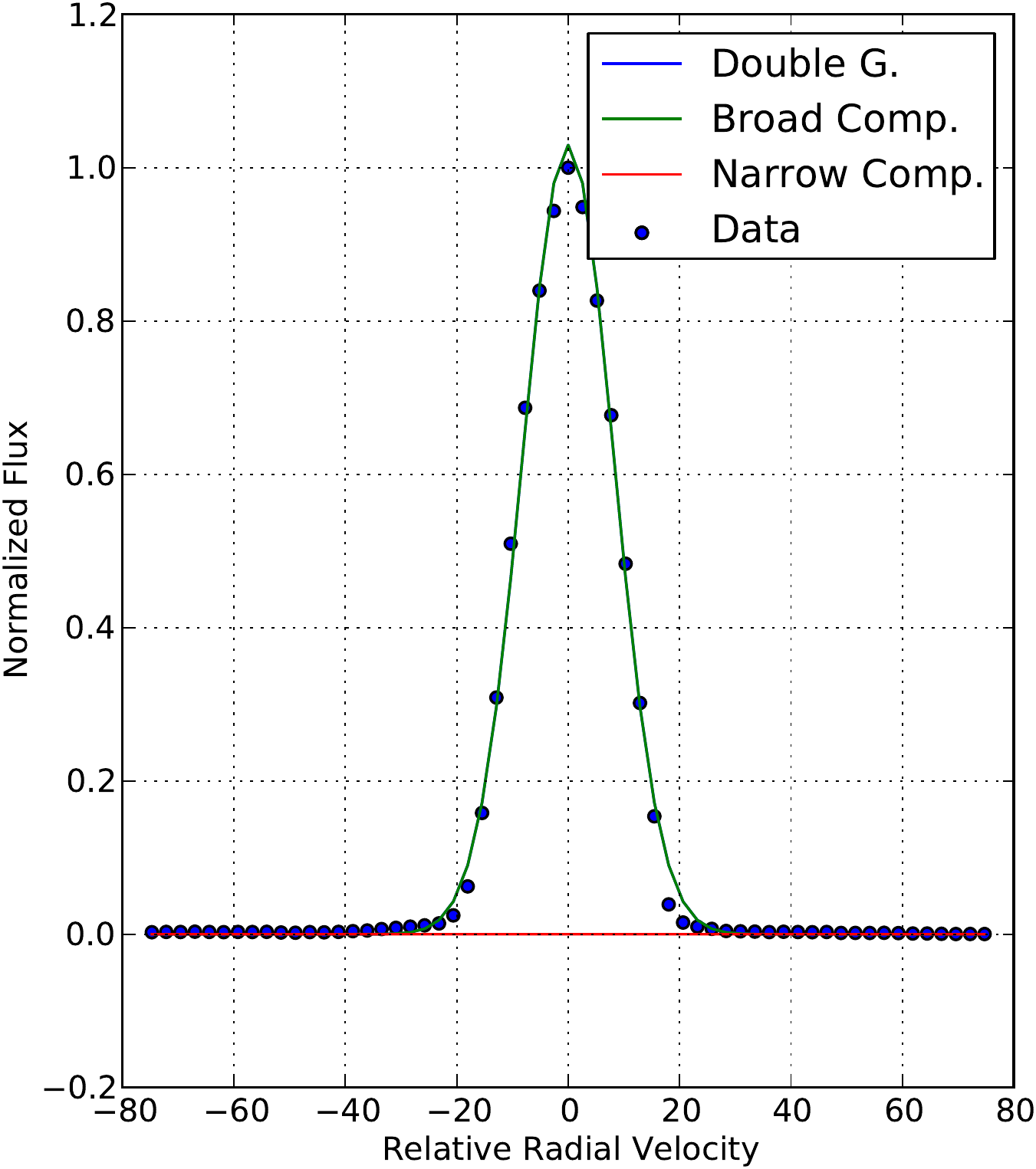}}
\subfigure[CHVC\,162+03-186 $\mathrm{Effelsberg_{\,reg}}$]{\includegraphics[scale=0.38]{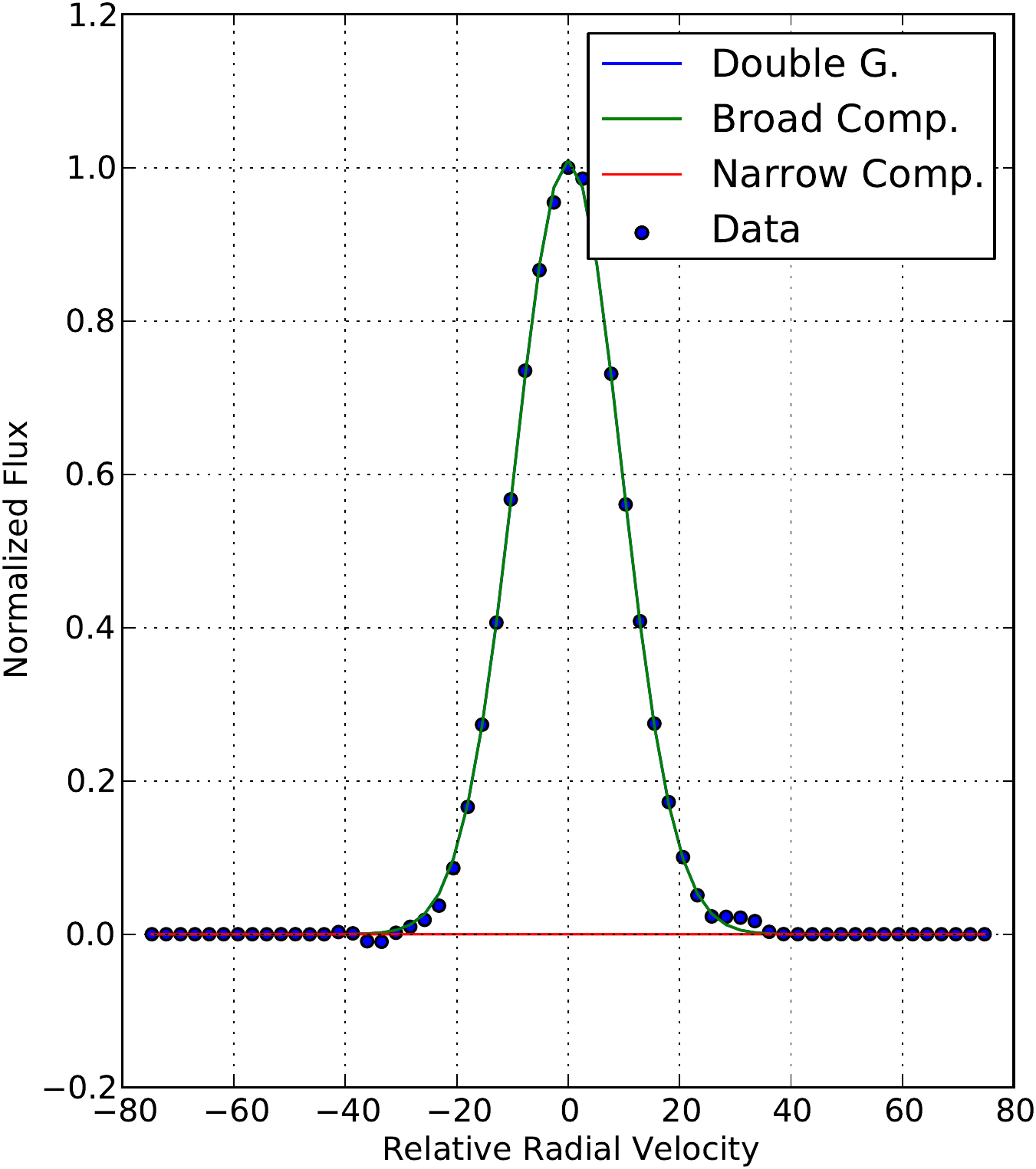}}
\caption{Super profiles of the three CHVCs. The first row presents the results of CHVC\,070+51-150. The middle row shows the results of  CHVC\,108-21-390 and the bottom row of CHVC\,162+03-186. The left panel displays the result of the Gaussian decomposition for the combined data. The middle and right panels present the results for the WSRT and Effelsberg data respectively. The green line corresponds to the broad component (WNM) and the red line to the narrow component (CNM).}
\label{fig:superprofiles}
\end{figure*}

Table\,\ref{tb:phy_par} summarizes the results of the Gaussian decomposition for all three CHVCs. The results of the Gaussian decomposition are also presented in Fig.\,\ref{fig:superprofiles}. The solid green line close to the data points (filled circles) traces the WNM while the small Gaussian in solid red represents the CNM.
The super profile \hi spectra have been renormalized. The x-axis presents the relative radial velocities in $\mathrm{km\,s^{-1}}$ in the super profile while the mean velocity is shifted to zero. The left panel (a) presents the results of the Gaussian decomposition for the combined data. The middle panel (b) and right panel (c) present the same values for the WSRT data and Effelsberg data respectively.

For all three CHVCs, we find only warm gas without any evidence of a cold neutral medium (see Fig.\ref{fig:superprofiles}). \citet{2006A&A...455..481K} detect rather broad lines for the ACVHVC, which agrees with the results of our analysis. Only in the case of CHVC\,108-21-390 is there marginal evidence of a second, cold gas component; yet it is intriguing to find no evidence of a CNM in any of the clouds. One possible interpretation could be that for these three clouds the interaction is not particularly strong owing to a combination of low velocity of the clouds relative to their environment and/or a low density of the ambient medium. In this case, the density within the core might not rise high enough to allow sufficient self-shielding and cooling of the gas. Some support for this could come from the morphology of the clouds: both CHVC\,070+51-150 and CHVC\,162+03-186 have their core relatively close to the center of the bulk of the \hi emission, rather than near the leading edge as one would expect for a strongly interacting cloud. The only exception is CHVC\,108-21-390. Its core is clearly shifted towards the presumed leading edge, and interestingly this is the only cloud that does indeed show some evidence of cold gas according to the Gaussian fits. 

In addition to the weak evidence of a CNM, CHVC\,108-21-390 has the narrowest lines of the three clouds. According to \citet{1995ApJ...453..673W}, a single component gas implies an upper limit for the density of $\approx 0.3 \, \mathrm{cm}^{-3}$. The single warm gas component derived from the analyses of the \hi super profiles implies Doppler temperatures of about $T_{\rm D} = \frac{m_H\,(\Delta v)^2}{8\,k\ln2}\,\simeq 21.8 \times (22\,{\rm km\,s^{-1}})^2 \simeq 10.000\,{\rm K}$. Using the derived Doppler temperature, we calculate an upper limit for the pressure of about $\approx 3300 {\rm K\,cm^{-3}}$, which is in agreement with the upper limit for a single-phase medium given by \citet{1995ApJ...453..673W} (their Fig.1, panel (e)). The parameter distance in the corresponding position in the phase diagram is completely degenerated, i.e., a cloud with the above parameters is always a simple, warm medium, regardless of distance. The physical parameters are also consistent with the ones from a cloud in thermal equilibrium with the ambient (halo) gas. This would support our proposal that the ram pressure interaction may play only a minor role because of the relatively slow movement of the cloud, as suggested before.

\section{Summary and outlook} \label{sec:conclusions}

We present deep integrated observations of the warm neutral medium for three CHVCs using the 100 m Effelsberg telescope and high-resolution observations of the more compact regions from the WSRT, as well as the results of their combination. Our approach shows that the combination in the image domain meets the expectations. The critical step in the algorithm is the regridding, where a flux inconsistency can occur as a result of interpolation inaccuracies. 

The combination results demonstrate the importance of the zero-spacing correction regarding determination of the physical and morphological properties of the objects. Here in particular the warm neutral medium is of major interest. High-velocity clouds in higher ambient pressure environments might show complex spatial and spectral structures. 

The results of the Gaussian decomposition suggest strongly that the three clouds mostly consist of WNM. The analysis of \citet{2011A&A...533A.105W} also confirms the lack of cold components in HVCs in the complex known as Galactic centre negative (GCN). However, high Doppler temperatures above $10^4$ K are an indication of the existence of turbulence within the clouds. The results demonstrate that the gas gets warmer or more turbulent in the tail region and at the edges of the clouds. This also suggests that the warm gas is floating away to the direction of the tail, which is an indication of existing ram pressure. Another remarkable aspect is the decreasing velocity gradient at the regions with lower column densities. The relatively slow movement of the clouds also reveals the lack of cold cores in the clouds, and the measured high Doppler temperatures demonstrate that the clouds are not in equilibrium. 

The linear approach used in the image domain is easy to implement and  is not very CPU-intensive. It could be very helpful for computation of huge amounts of data coming from telescopes like ASKAP and WSRT/APERTIF.

\begin{acknowledgements}
The authors are grateful to the Deutsche Forschungsgemeinschaft (DFG) for support under grant numbers KE757/7-1, KE757/7-2 \& KE757/9-1.

Based on observation performed by the 100 m Effelsberg telescope operated by Max-Planck-Institut f\"ur Radioastronomie (MPIfR) in Bonn. Based on observations with the Westerbork Synthesis Radio Telescope (WSRT) operated by ASTRON in Dwingeloo. 

The lead author is very thankful to Tobias R\"ohser, Ian Stewart from Argelander-Institut f\"ur Astronomie (AIfA) and the referee, B. Wakker, for their very useful comments and suggestions.
\end{acknowledgements}

\bibliographystyle{aa}    
\bibliographystyle{natbib}    
\bibliography{chvcs}        

\end{document}